# Influence of surface integrity on geometry and dynamics of functionally graded nanobeams

## Mohamed Shaat*

*Engineering and Manufacturing Technologies Department, DACC, New Mexico State University, Las Cruces, NM 88003, USA*

*Mechanical Engineering Department, Zagazig University, Zagazig 44511, Egypt*

**Abstract**

In this study effects of surface integrity (*i.e.* surface texture and surface mechanical properties) on the mechanics of functionally graded (FG) nanobeams are investigated. This study reports the changes in the geometry and dynamics of FG nanobeams because of changes in their surface textures and/or surface mechanical properties.

A new model for FG nanobeams with engineering surfaces is developed. This engineering surface is considered as a different material phase with a surface texture (*i.e.* waviness and roughness). The initial curvatures of cantilever, simple supported, and clamped-clamped FG nanobeams due to surface residual stresses are determined. Moreover, their natural frequencies and mode shapes are derived depending on the surface integrity.

The initial curvatures of FG beams are obtained increasing with an increase in the slope of the surface texture and/or a decrease in the heights of the asperities of the surface roughness. Moreover, it is observed that the natural frequencies of FG beams may decrease or increase due surface integrity depending on the boundary conditions. Thus, as a first prospect, the surface roughness allows the vibration energy to propagation over the beam length and hence its natural frequency decreases resulting in a zero-frequency mode. As for the other prospect, surface roughness inhibits the propagation of the vibration energy through the beam length leading to a mode localization. It is revealed that a mode localization is accompanied with an increase in the natural frequency of the nanobeam.

The proposed surface integrity model for FG nanobeams is compared with Gurtin-Murdoch surface elasticity model. The results demonstrate that the surface integrity model is preferred over the former model where it accounts for, both, surface texture and surface mechanical properties effects. However, Gurtin-Murdoch model assumes smooth surfaces of nanobeams which leads to under/overestimations of their mechanics.

**Keywords**: vibration; mode localization; zero-frequency mode, roughness; nanobeams.

## 1. Introduction

A functionally graded (FG) material is an inhomogeneous composite material in which the volume fractions of the material constituents vary smoothly and continuously through the microstructure. In 1984, the concept of functionally graded materials was introduced [Koizumi, 1997] to announce the initiation of a new era of advanced materials. FG materials allow engineers control their microstructures to develop designs of smart and multi-functional systems. Therefore, FG materials have been implemented in many engineering applications. For example, due to the high thermal stability of metallic-ceramic FG materials, they are being used as thermal coatings for barrier engines, gas turbines, and nuclear fusions [Alshorbagy et al., 2013]. Also, FG materials are being used for making optical thin layers and biomaterial electronics [Shaat et al., 2012].

---------------------------------------------------------------------------------------------------

*Corresponding author. Tel.: +15756215929.
E-mail addresses: shaat@nmsu.edu; shaatscience@yahoo.com (M. Shaat).

Nowadays, micro/nano-sized structures made of FG materials are being proposed to design smart structures for various engineering and medical applications. Therefore, studies were carried out to investigate the mechanics of FG micro/nanostructures such as bars, beams, plates, and shells. Moreover, because their mechanics is affected with the excess energy of their surfaces, a countless list of studies used Gurtin-Murdoch model of the surface elasticity [*e.g.* Lü et al., 2009a, 2009b; Shaat et al., 2012, 2013a, 2013b, 2013c; Mahmoud and Shaat, 2015; Alshorbagy et al., 2013; Sharabiani and Yazdi, 2013; Ansari et al., 2015; Hosseini-Hashemi et al., 2014; Hosseini-Hashemi and Nazemnezhad, 2013] to examine bending, vibration, and buckling of FG nanostructures under mechanical and thermal loads. For instance, using Gurtin-Murdoch surface elasticity model, Lü et al. [2009a, 2009b], Shaat et al. [2012, 2013a, 2013b], and Mahmoud and Shaat [2015] studied the static bending of FG ultra-thin films. Alshorbagy et al. [2013] developed a finite element model for bending of FG plates exposed to thermomechanical loads. Sharabiani and Yazdi [2013] studied the free vibration of FG nanobeams. Ansari et al. [2015] investigated the vibration and buckling characteristics of FG nanoplates under thermal loads. Hosseini-Hashemi et al. [2014] investigated the effects of the surface energy and the material nonlocality on free vibration of piezoelectric FG nanobeams. Hosseini-Hashemi and Nazemnezhad [2013] studied the influence of the geometrical nonlinearity and the surface energy effects on the vibration of FG nanobeams.

Due to uncontrollable manufacturing errors, engineering structures may possess material and/or geometrical irregularities. The presence of these irregularities in a structure may result in significant changes in its dynamical behaviors. For instance, due to a small irregularity in periodic structures, a mode localization or an eigenvalue loci veering may occur [Kim and Lee, 1998; Chan and Liu, 2000; Hodges, 1982; Bendiksen, 1987; Pierre, 1988; Cai and Lin, 1991; Triantafyllou and Triantafyllou, 1991; Natsiavas, 1993; Xie, 1995; Cai et al., 1995; Liu et al., 1995; Liu et al., 1996; Pierre et al., 1996]. Mode localization is one of the common changes in the normal modes of structures due to irregularities where the vibration energy is confined in a specific portion of the structure. Eigenvalue loci veering is that two successive frequencies approach each other and then veer apart with a high local curvature [Chan and Liu, 2000; Pierre, 1988; Liu et al., 1995; Sari et al., 2017]. Eigenvalue loci veering usually causes a mode shape change [Sari et al. 2017] or a mode localization [Pierre, 1988].

Small-scale structures are more sensitive to irregularities than large-scale structures. Mode localization and eigenvalue loci veering phenomena in micro/nanobeam-structures with material and/or geometrical irregularities were discussed in a very short list of studies: Dehrouyeh-Semnani et al. [2016] discussed the mode veering phenomenon in curved FG microbeams, Sari et al. [2017] revealed frequency and mode veering phenomena in axially-FG non-uniform beams, and Pradiptya and Ouakad [2017] demonstrated that when a buckled nanobeam is exposed to a high temperature, mode veering may take place. Indeed, more studies on exploring the mode localization and eigenvalue loci veering phenomena in micro/nanostructure should be carried out.

As previously mentioned, Gurtin-Murdoch surface elasticity model was intensively used to model the mechanics of nanostructures. Gurtin-Murdoch model introduces measures to account for the influence of the excess energy of the material surface on its mechanics. However, this former model assumes a nominal surface of the material. An engineering surface, on the other hand, contains various forms of irregularities, *e.g.* waviness and roughness.

Surface integrity is the characterization of an engineering surface which concerns with describing the surface texture, metallurgy, and mechanical properties [Shaat, 2017a; Bellows and Tishler, 1970; Astakhov 2010]. A surface texture is the representation of the outermost layer of the material which includes irregularities such as waviness, roughness, faults, and cracks. The metallurgy of the surface is the representation of the microstructure of the altered layer which is confined between the outermost layer and the material bulk. The great impact the surface integrity on the mechanical properties and performance of materials has been demonstrated in various studies. For example, effects of the surface integrity on the fatigue properties of metals [Sharman et al., 2001; Dieter, 1988; Zaltin and Field, 1973; Huang and Ren, 1991; Ramulu et al., 2001; Sinnott et al., 1989], the fracture resistance of cemented carbides [Llanes et al., 2004], the adhesion between elastic bodies [Peressadko et al., 2005, Greenwood and Williamson, 1966], the wettability of hydrophobic surfaces, and the conductivity [Fishman and Calecki, 1991] and magnetic properties [Li et al., 1998] of metals were investigated.

Recently, Shaat [2017] proposed the surface integrity model. The surface integrity model outweighs Gurtin-Murdoch surface model [Gurtin and Murdoch, 1975a, 1975b, 1978; Murdoch, 2005] by modeling a material with an engineering surface. Shaat's model of surface integrity introduces measures to account for the influence of the surface integrity (including waviness, roughness, altered layers, and surface excess energy) on the mechanics of micro/nano-sized materials. In this study, effects of the surface integrity on the vibration characteristics of FG nanobeams are explored. Thus, the frequencies and mode shapes of FG nanobeams with various boundary conditions are determined based on the surface integrity model.

This effort presents the first study on the effects of surface integrity on mechanics of FG nanobeams. In this study, the changes in the geometry and dynamics of FG nanobeams because of changes in their surface textures and/or surface mechanical properties are reported. To this end, a surface integrity model for FG nanobeams with engineering surfaces is developed. The surface is considered with a surface texture (*i.e.* waviness and roughness). The proposed surface integrity model is formulated in section 2 where two versions of the model are developed. In the first version, the surface integrity model of FG beams is derived depending on a surface profile function. Then, the model is reformulated depending of the average parameters of the surface texture. The initial curvatures of cantilever, simple supported, and clamped-clamped FG nanobeams due to surface residual stresses are determined in section 3. Moreover, their natural frequencies and mode shapes are derived in section 4. A parametric study on effects of the surface integrity on the initial curvature, frequencies, and mode shapes of FG nanobeams is provided in section 5. Moreover, a comparison between the proposed surface integrity model for FG nanobeams and Gurtin-Murdoch surface elasticity model and the classical model is presented in section 6. The conclusions and the findings of this study are summarized in section 7.

## 2. Formulation of surface integrity effects on mechanics of FG nanobeams

In this section, a detailed formulation for the influence of the surface integrity on the mechanics of FG nanobeams is developed. This formulation is inspired by the surface integrity model [Shaat, 2017a]. Thus, a FG nanobeam is modeled as a material bulk, $\Omega_B$, with an engineering surface, $S$. This engineering surface reflects the real surface texture including surface waviness and surface roughness of the FG beam. Moreover, the engineering surface is modeled with different mechanical properties and with a residual surface stress.

In this study, a FG beam is modeled as a linear elastic-Euler-Bernoulli beam; therefore, the displacement field, $u_i$, of a point, $\mathbf{x} = (x, y, z)$, that belongs to the beam can be defined as follows:

$$u_x(x, z, t) = -z \frac{\partial w(x,t)}{\partial x}, u_y = 0, \ u_z(x, t) = w(x, t) \tag{1}$$

where $w(x, t)$ denotes the transverse deflection of the beam. Also, the non-zero component of the infinitesimal strain, $\varepsilon_{ij} = \frac{1}{2}(u_{i,j} + u_{j,i})$ *i.e.* $i, j \equiv x, y, z$, can be obtained as follows:

$$\varepsilon_{xx}(x, z, t) = -z \frac{\partial^2 w(x, t)}{\partial x^2} \tag{2}$$

To model the surface texture of a FG beam, $P(x)$ is introduced as the profile of the outermost surface layer of the beam:

$$P(x) = \mathcal{R}(x) + \varpi(x) \tag{3}$$

where $\mathcal{R}(x)$ and $\varpi(x)$ denote the profiles of the surface roughness and waviness, respectively. It should be mentioned that $\mathcal{R}(x)$ represents the deviations of the asperities of the surface roughness measured from the waviness, as shown in Figure 1.

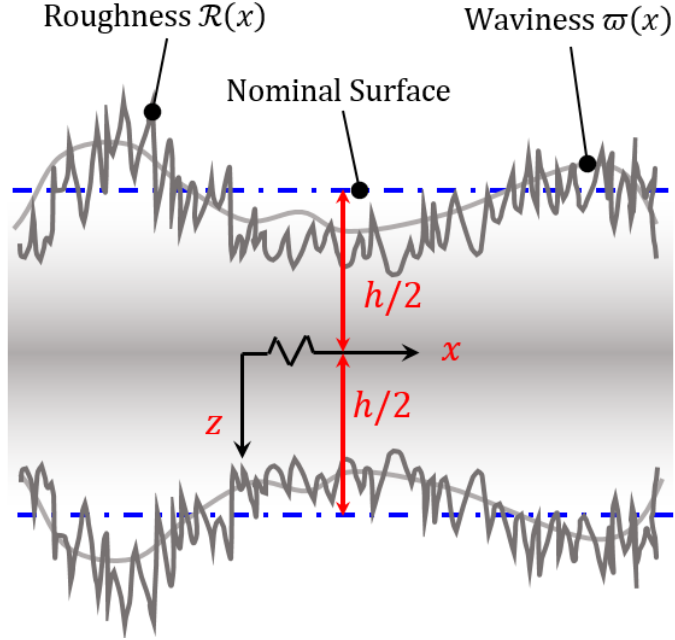

Figure 1: A schematic of a FG beam with an engineering surface.

By considering an isotropic FG beam with a material variation along z-direction, the non-zero stress components of the material bulk, $\Omega_B$, are obtained as follows:

$$\sigma_{xx}(x,z,t) = E^*(x,z)\varepsilon_{xx}(x,z,t) \ , \ \sigma_{yy} = \sigma_{zz} = \lambda(x,z)\varepsilon_{xx} \qquad \forall (x,y,z) \in \Omega_B \qquad (4)$$

where $\sigma_{ij}$ denote the components of the stress tensor of the material bulk. $E^*$ is the equivalent elastic modulus of the beam interior (material bulk). $\lambda$ is the beam's 1st Lame constant.

It should be mentioned that the material properties of the beam depend on the material variation through the beam thickness as well as the geometrical variations due to the surface texture.

Because it may be exposed to a high temperature during processing and because of the environmental interactions, the surface of the beam possesses different mechanical properties. Thus, the surface of the beam experiences a different strain energy level than the material bulk. Furthermore, because atoms at the free surface of the beam are under-coordinated, a residual stress is generated to achieve the equilibrium at the surface. Given these observations, the total stress at the beam surface can be defined as follows:

$$\sigma_{xx}^S(x,z,t) = \sigma_S(x,z) + E_S^*(x,z)\varepsilon_{xx}(x,z,t) \ ,$$
$$\sigma_{yy}^S = \sigma_{zz}^S = \lambda_S(x,z)\varepsilon_{xx} \qquad\qquad \forall (x,y,z) \in S \qquad (5)$$

where $\sigma_{ij}^S$ denote the components of the stress at the outermost surface of the beam. $E_S^*$ is the equivalent elastic modulus of the beam surface. $\sigma_S$ denotes the residual surface stress.

To derive the equations of motion, Hamilton's principle is employed which states:

$$\int\limits_0^t (\delta T - \delta U + \delta Q)\, dt = 0 \tag{6}$$

where $\delta U$, $\delta T$, and $\delta Q$ are the first variations of the strain energy, the kinetic energy, and the work done, respectively. According to the considered formulation, these quantities can be defined as follows:

$$\delta T = -\int\limits_{\Omega_B} \rho(x,z)\ddot{w}\delta w\, d\Omega_B \tag{7}$$

$$\delta U = \int\limits_{\Omega_B} \sigma_{xx}^B \delta\varepsilon_{xx}\, d\Omega_B + \int\limits_S \sigma_{xx}^S \delta\varepsilon_{xx}\, dS \tag{8}$$

$$\delta Q = \int\limits_{\Omega_B} (f_z(x,t)\delta u_z)d\Omega_B + \oint\limits_\Gamma (t_z\delta u_z)dS \tag{9}$$

where $\rho$ is the beam mass density. $f_z(x,t)$ denotes the body force in the z-direction. $t_z$ is the surface traction along z-direction.

By substituting equations (7)-(9) into equation (6), the equations of motion of a FG beam modeled based on the surface integrity model can be obtained in the following form:

$$\frac{\partial^2 M_{xx}(x,t)}{\partial x^2} + F_z(x,t) - I_0(x)\frac{\partial^2 w(x,t)}{\partial t^2} = 0 \tag{10}$$

with

$$F_z(x,t) = \int\limits_{A(x)} f_z(x,t)\, dA$$

$$I_0(x) = \int\limits_{A(x)} \rho(x,z)\, dA \tag{11}$$

where $A(x)$ is the non-uniform cross sectional area of the beam due to its surface texture. $I_0(x)$ denotes the mass per unit length.

The corresponding boundary conditions are obtained as follows:

$$\frac{\partial M_{xx}}{\partial x} = \bar{V} \quad \text{or} \quad w = w^\Gamma$$

$$M_{xx} = \bar{M} \quad \text{or} \quad \frac{\partial w}{\partial x} = w_{,x}^\Gamma \tag{12}$$

where $\bar{V}$ and $\bar{M}$ denote the applied shear force and bending moment at the beam end, $\Gamma$. $w^\Gamma$ and $w_{,x}^\Gamma$ are prescribed deflection and slope at the beam end.

The moment stress resultant introduced in equations (10) is obtained for the considered FG beams as follows:

$$M_{xx}(x,t) = \int_{A_B} z\sigma_{xx}^B \, dA + \int_S z\sigma_{xx}^S \, dS = -D(x)\frac{\partial^2 w(x,t)}{\partial x^2} + q(x) \tag{13}$$

where

$$D(x) = D_B(x) + D_S(x) \qquad \text{and} \quad q(x) = \int_S z\sigma_S(x,z)\,dS$$

with

$$D_B(x) = \int_{A_B} z^2 E^*(x,z)\,dA \tag{14}$$

$$D_S(x) = \int_S z^2 E_S^*(x,z)\,dS$$

By substituting equation (13) into equation (10), the equation of motion of a FG beam according to the surface integrity model can be obtained in terms of the beam deflection as follows:

$$D(x)\frac{\partial^4 w(x,t)}{\partial x^4} + K(x)\frac{\partial^3 w(x,t)}{\partial x^3} + B(x)\frac{\partial^2 w(x,t)}{\partial x^2} - \frac{\partial^2 q(x)}{\partial x^2} - F_z(x,t)$$
$$+ I_0(x)\frac{\partial^2 w(x,t)}{\partial t^2} = 0 \tag{15}$$

where

$$K(x) = 2\frac{dD(x)}{dx}$$
$$B(x) = \frac{d^2 D(x)}{dx^2} \tag{16}$$

Similarly, by substituting equation (13) into equation (12), the boundary conditions can be written in terms of the beam deflection as follows:

$$-D(x)\frac{\partial^3 w(x,t)}{\partial x^3} + \frac{\partial q(x)}{\partial x} - \frac{1}{2}K(x)\frac{\partial^2 w(x,t)}{\partial x^2} = \bar{V} \quad \text{or} \quad w = w^\Gamma$$
$$-D(x)\frac{\partial^2 w(x,t)}{\partial x^2} + q(x) = \bar{M} \quad \text{or} \quad \frac{\partial w}{\partial x} = w_{,x}^\Gamma \tag{17}$$

For a FG beam with an engineering surface whose material varies according to the simple power law through the beam thickness, its material properties can be defined as follows:

$$E(x,z) = E^+ + (E^- - E^+)\left(\frac{z}{2\left(\frac{h}{2}+P(x)\right)} + \frac{1}{2}\right)^n \quad i.e. \ n \geq 0$$

$$E_S(x,z) = E_S^+ + (E_S^- - E_S^+)\left(\frac{z}{2\left(\frac{h}{2}+P(x)\right)} + \frac{1}{2}\right)^n \quad i.e. \ n \geq 0 \tag{18}$$

$$\sigma_S(x,z) = \sigma_S^+ + (\sigma_S^- - \sigma_S^+)\left(\frac{z}{2\left(\frac{h}{2}+P(x)\right)}+\frac{1}{2}\right)^n \quad i.e.\ n \geq 0$$

where $h$ is the beam thickness. The superscript '+' refers to the material properties at the upper surface of the beam ($z = -h/2 - P(x)$). The subscript '−' refers to the material properties at the lower surface of the beam ($z = h/2 + P(x)$).

By substituting equation (18) into equation (14) and performing the integration:

$$D_B(x) = 2\left(\frac{b}{2}+P(x)\right)\left(\frac{h}{2}+P(x)\right)^3\left(\frac{2}{3}E^+ + (E^- - E^+)\xi(n)\right)$$

$$D_S(x) = 2\left(\frac{b}{2}+P(x)\right)\left(\frac{h}{2}+P(x)\right)^2(E_S^+ + E_S^-) + 2\left(\frac{h}{2}+P(x)\right)^3\left(\frac{2}{3}E_S^+ + (E_S^- - E_S^+)\xi(n)\right) \quad (19)$$

$$q(x) = 2\left(\frac{b}{2}+P(x)\right)\left(\frac{h}{2}+P(x)\right)(\sigma_S^- - \sigma_S^+) + 2\left(\frac{h}{2}+P(x)\right)^2(\sigma_S^- - \sigma_S^+)\kappa(n)$$

where $\xi(n)$ and $\kappa(n)$ are two hypergeometric functions only depend on the grading parameter, $n$. These functions are defined as follows:

$$\xi(n) = \frac{2^{-n}}{n+3}[\text{hypergeom}([-n,-n-3],[-n,-2],-1)$$
$$+ (-1)^n\text{hypergeom}([-n,-n-3],[-n,-2],1)] \quad (20)$$
$$\kappa(n) = \frac{(-1)^n\text{hypergeom}([-n-2],[\ \ ],1)}{2^n(n+2)(n+1)}[-2n-3] + \frac{2n}{(n+2)(n+1)}$$

According to equation (19), the stiffnesses, $K(x)$ and $B(x)$ can be derived as follows:

$$K(x) = 4\left(\frac{2}{3}E^+ + (E^- - E^+)\xi(n)\right)\left(\frac{dP(x)}{dx}\right)\left[3\left(\frac{b}{2}+P(x)\right)\left(\frac{h}{2}+P(x)\right)^2 + \left(\frac{h}{2}+P(x)\right)^3\right]$$

$$+ 4(E_S^+ + E_S^-)\left(\frac{dP(x)}{dx}\right)\left[2\left(\frac{b}{2}+P(x)\right)\left(\frac{h}{2}+P(x)\right) + \left(\frac{h}{2}+P(x)\right)^2\right]$$

$$+ 12\left(\frac{2}{3}E_S^+ + (E_S^- - E_S^+)\xi(n)\right)\left(\frac{dP(x)}{dx}\right)\left(\frac{h}{2}+P(x)\right)^2$$

$$B(x) = 6\left(\frac{2}{3}E^+ + (E^- - E^+)\xi(n)\right)\left(\frac{dP(x)}{dx}\right)^2\left[\left(\frac{h}{2}+P(x)\right)^2 + 2\left(\frac{b}{2}+P(x)\right)\left(\frac{h}{2}+P(x)\right)\right] \quad (21)$$

$$+ \left(\frac{h}{2}+P(x)\right)^2\right] + 4(E_S^+ + E_S^-)\left(\frac{dP(x)}{dx}\right)^2\left[2\left(\frac{h}{2}+P(x)\right) + \left(\frac{b}{2}+P(x)\right)\right]$$

$$+ 12\left(\frac{2}{3}E_S^+ + (E_S^- - E_S^+)\xi(n)\right)\left(\frac{dP(x)}{dx}\right)^2\left(\frac{h}{2}+P(x)\right)$$

In the derived model, $P(x)$ is incorporated to account for the influence of surface integrity on the mechanics of FG beams. As demonstrated by the author in [Shaat, 2017a], it is challenging to derive analytical solutions for the equation of motion (equation (15)) utilizing the profile function, $P(x)$. Moreover, the surface integrity as reported by the experiment is explained with the average of the surface parameters [Shaat, 2017b]. Therefore, following the approach proposed by the author in [Shaat, 2017a], the derived model can be rewritten considering the average of the surface profile and the average of its slope, as follows:

$$\langle P(x) \rangle = W_a + R_a$$
$$\langle dP(x)/dx \rangle = WS + RS \tag{22}$$

where $R_a$ is the average roughness, and $W_a$ is the average waviness. $RS$ denotes the average slope of the surface roughness, and $WS$ is the average slope of the surface waviness. These average parameters can be defined for a FG beam as follows:

$$R_a = \frac{1}{L} \int_0^L |\mathcal{R}(x)| \, dx$$

$$W_a = \frac{1}{L} \int_0^L |\varpi(x)| \, dx$$

$$RS = \frac{1}{L} \int_0^L \left| \frac{d\mathcal{R}(x)}{dx} \right| dx \tag{23}$$

$$WS = \frac{1}{L} \int_0^L \left| \frac{d\varpi(x)}{dx} \right| dx$$

where $L$ is the beam length.

Utilizing the average parameters defined in equation (23), the equation of motion (15) and the boundary conditions (17) can be rewritten as follows:

$$D \frac{\partial^4 w(x,t)}{\partial x^4} + K \frac{\partial^3 w(x,t)}{\partial x^3} + B \frac{\partial^2 w(x,t)}{\partial x^2} - S_f - F_z(x,t) + I_0 \frac{\partial^2 w(x,t)}{\partial t^2} = 0 \tag{24}$$

$$-D \frac{\partial^3 w(x,t)}{\partial x^3} + Q - \frac{1}{2} K \frac{\partial^2 w(x,t)}{\partial x^2} = \bar{V} \quad \text{or} \quad w = w^\Gamma$$

$$-D \frac{\partial^2 w(x,t)}{\partial x^2} + q = \bar{M} \quad \text{or} \quad \frac{\partial w}{\partial x} = w_{,x}^\Gamma \tag{25}$$

where

$$D = 2\left(\frac{b}{2} + W_a + R_a\right)\left(\frac{h}{2} + W_a + R_a\right)^3\left(\frac{2}{3}E^+ + (E^- - E^+)\xi(n)\right)$$
$$+ 2\left(\frac{b}{2} + W_a + R_a\right)\left(\frac{h}{2} + W_a + R_a\right)^2(E_S^+ + E_S^-)$$
$$+ 2\left(\frac{h}{2} + W_a + R_a\right)^3\left(\frac{2}{3}E_S^+ + (E_S^- - E_S^+)\xi(n)\right) \tag{26}$$

$$K = 4\left(\frac{2}{3}E^+ + (E^- - E^+)\xi(n)\right)(WS + RS)\left[3\left(\frac{b}{2} + W_a + R_a\right)\left(\frac{h}{2} + W_a + R_a\right)^2\right.$$
$$\left.+ \left(\frac{h}{2} + W_a + R_a\right)^3\right]$$
$$+ 4(E_S^+ + E_S^-)(WS + RS)\left[2\left(\frac{b}{2} + W_a + R_a\right)\left(\frac{h}{2} + W_a + R_a\right)\right.$$
$$\left.+ \left(\frac{h}{2} + W_a + R_a\right)^2\right]$$
$$+ 12\left(\frac{2}{3}E_S^+ + (E_S^- - E_S^+)\xi(n)\right)(WS + RS)\left(\frac{h}{2} + W_a + R_a\right)^2 \tag{27}$$

$$B = 6\left(\frac{2}{3}E^+ + (E^- - E^+)\xi(n)\right)(WS + RS)^2\left[\left(\frac{h}{2} + W_a + R_a\right)^2\right.$$
$$\left.+ 2\left(\frac{b}{2} + W_a + R_a\right)\left(\frac{h}{2} + W_a + R_a\right) + \left(\frac{h}{2} + W_a + R_a\right)^2\right]$$
$$+ 4(E_S^+ + E_S^-)(WS + RS)^2\left[2\left(\frac{h}{2} + W_a + R_a\right) + \left(\frac{b}{2} + W_a + R_a\right)\right]$$
$$+ 12\left(\frac{2}{3}E_S^+ + (E_S^- - E_S^+)\xi(n)\right)(WS + RS)^2\left(\frac{h}{2} + W_a + R_a\right) \tag{28}$$

$$q = 2\left(\frac{b}{2} + R_a + W_a\right)\left(\frac{h}{2} + R_a + W_a\right)(\sigma_S^- - \sigma_S^+) + 2\left(\frac{h}{2} + R_a + W_a\right)^2(\sigma_S^- - \sigma_S^+)\kappa(n) \tag{29}$$

$$Q = 2(\sigma_S^- - \sigma_S^+)(WS + RS)\left[\left(\frac{b}{2} + W_a + R_a\right) + \left(\frac{h}{2} + W_a + R_a\right)\right]$$
$$+ 4(\sigma_S^- - \sigma_S^+)(WS + RS)\left(\frac{h}{2} + W_a + R_a\right)\kappa(n) \tag{30}$$

$$S_f = 4(\sigma_S^- - \sigma_S^+)(WS + RS)^2 + 4(\sigma_S^- - \sigma_S^+)(WS + RS)^2\kappa(n) \tag{31}$$

where $S_f$ is a residual force due to surface stresses.

It should be mentioned that the higher-order gradients of the surface profile, $P(x)$, are neglected as recommended by the author in [Shaat, 2017a].

Equations (15)-(21) represent a model of FG beams with engineering surfaces according to the surface integrity model. The formulation presented in these equations depends on the surface profile function, $P(x)$.

This model is reformulated considering the average of the surface profile as presented in equations (24)-(31).

The derived model recovers the model of FG beams based on Gurtin-Murdoch surface elasticity model by setting $R_a \to 0$, $W_a \to 0$, $RS \to 0$, and $WS \to 0$ in the derived equations (24)-(31), as follows:

$$D \frac{\partial^4 w(x,t)}{\partial x^4} - F_z(x,t) + I_0 \frac{\partial^2 w(x,t)}{\partial t^2} = 0 \tag{32}$$

$$-D \frac{\partial^3 w(x,t)}{\partial x^3} = \bar{V} \quad \text{or} \quad w = w^\Gamma$$

$$-D \frac{\partial^2 w(x,t)}{\partial x^2} + q = \bar{M} \quad \text{or} \quad \frac{\partial w}{\partial x} = w_{,x}^\Gamma \tag{33}$$

where

$$D = \left(\frac{bh^3}{8}\right)\left(\frac{2}{3}E^+ + (E^- - E^+)\xi(n)\right) + \left(\frac{bh^2}{4}\right)(E_S^+ + E_S^-) + \left(\frac{h^3}{4}\right)\left(\frac{2}{3}E_S^+ + (E_S^- - E_S^+)\xi(n)\right) \tag{34}$$

$$q = \left(\frac{bh}{2}\right)(\sigma_S^- - \sigma_S^+) + \left(\frac{h^2}{2}\right)(\sigma_S^- - \sigma_S^+)\kappa(n) \tag{35}$$

It is clear that the surface integrity model outweighs Gurtin-Murdoch model by modeling the real engineering surface of the material. Thus, the former model assumes a smooth surface of the material. The surface integrity model, on the other hand, accounts for, both, the surface texture and the surface excess energy effects on the mechanics of nano-sized materials.

The nondimensional form of the derived equations (24) and (25) can be obtained by using the following nondimensional parameters:

$$W(X) = \frac{\sqrt{12}w(x)}{h}, X = \frac{x}{L}, T = t\sqrt{\frac{D}{I_0 L^4}} \tag{36}$$

The substitution of equation (36) into equations (24) and (25) gives:

$$\frac{\partial^4 W(X,T)}{\partial X^4} + \alpha \frac{\partial^3 W(X,T)}{\partial X^3} + \beta \frac{\partial^2 W(X,T)}{\partial X^2} + \frac{\partial^2 W(X,T)}{\partial T^2} = \bar{F}_z(X,T) + \bar{S}_f \tag{37}$$

$$\frac{\partial^3 W(X,T)}{\partial X^3} + \frac{1}{2}\alpha \frac{\partial^2 W(X,T)}{\partial X^2} = -\bar{\aleph} + \frac{QL^3\sqrt{12}}{Dh} \quad \text{or } W = W^\Gamma$$

$$\frac{\partial^2 W(X,T)}{\partial X^2} = -\bar{\mathcal{M}} + \frac{qL^2\sqrt{12}}{Dh} \quad \text{or } \frac{\partial W(X,T)}{\partial X} = W_{,X}^\Gamma \tag{38}$$

with

$$\alpha = \frac{KL}{D}, \beta = \frac{BL^2}{D}, \bar{S}_f = \frac{S_f L^4 \sqrt{12}}{Dh}$$

$$\bar{F}_z(X,T) = \frac{F_z(x,t)L^4\sqrt{12}}{Dh} \tag{39}$$

where $\bar{\aleph}$ and $\bar{\mathcal{M}}$ are the nondimensional applied shear force and bending moment. $W^\Gamma$ and $W_{,X}^\Gamma$ are the nondimensional prescribed deflection and slope.

## 3. Static analysis: initial curved configurations of FG nanobeams due to surface stresses

In this section, the initial curved configurations of FG beams due to surface residual stresses are analytically derived. To this end, equations (37) and (38) are rewritten after dropping the external force-terms and the time-dependent term as follows:

$$\frac{\partial^4 W_s(X)}{\partial X^4} + \alpha \frac{\partial^3 W_s(X)}{\partial X^3} + \beta \frac{\partial^2 W_s(X)}{\partial X^2} = \bar{S}_f \qquad (40)$$

$$\frac{\partial^3 W_s(X)}{\partial X^3} + \frac{1}{2}\alpha \frac{\partial^2 W_s(X)}{\partial X^2} = \frac{QL^3\sqrt{12}}{Dh} \qquad \text{or } W_s = W_s^\Gamma$$

$$\frac{\partial^2 W_s(X)}{\partial X^2} = \frac{qL^2\sqrt{12}}{Dh} \qquad \text{or } \frac{\partial W_s(X)}{\partial X} = W_{s,X}^\Gamma \qquad (41)$$

Where $W_s$ denotes the initial curvature of a FG beam.

A general solution of equation (40) is obtained in the form:

$$W_s(X) = a_1 + a_2 x + \exp(-\alpha x/2)\,(a_3 \cos(\lambda_s x) + a_4 \sin(\lambda_s x)) \qquad (42)$$

where $\lambda_s = \frac{1}{2}\sqrt{4\beta - \alpha^2}$.

In this study, the initial curvatures of cantilever, simple supported, and clamped-clamped FG nanobeams are determined. Therefore, the following boundary conditions are employed:

Cantilever:

$$W_s(0) = 0, \frac{\partial W_s(0)}{\partial X} = 0, \frac{\partial^2 W_s(1)}{\partial X^2} = \frac{qL^2\sqrt{12}}{Dh}, \frac{\partial^3 W_s(1)}{\partial X^3} = \frac{QL^3\sqrt{12}}{Dh} - \frac{1}{2}\alpha\left(\frac{qL^2\sqrt{12}}{Dh}\right) \qquad (43)$$

Simple supported:

$$W_s(0) = 0, \frac{\partial^2 W_s(0)}{\partial X^2} = \frac{qL^2\sqrt{12}}{Dh}, W_s(1) = 0, \frac{\partial^2 W_s(1)}{\partial X^2} = \frac{qL^2\sqrt{12}}{Dh} \qquad (44)$$

Clamped-clamped:

$$W_s(0) = 0, \frac{\partial W_s(0)}{\partial X} = 0, W_s(1) = 0, \frac{\partial W_s(1)}{\partial X} = 0 \qquad (45)$$

Thus, the constants, $a_1$, $a_2$, $a_3$, and $a_4$, are determined for the different FG beams by applying the boundary conditions defined in equations (43)-(45).

## 4. Eigenvalue problem analysis of a curved FG beam

In this section, the free vibrations about the initial curved configurations of cantilever, simple supported, and clamped-clamped FG nanobeams are determined. Therefore, the beam deflection is decomposed as follows:

$$W(X,T) = W_s(X) + \varphi(X)\exp(i\omega T) \qquad (46)$$

where $\omega$ is the nondimensional natural frequency, and $\varphi(X)$ is the corresponding mode shape.

By substituting equation (46) into equations (37) and (38), substituting equations (40) and (41) into the result and dropping the forcing terms, the equations governing the mode shape of the free vibration of FG beams according to the surface integrity model are obtained as follows:

$$\frac{d^4\varphi(X)}{dX^4} + \alpha\frac{d^3\varphi(X)}{dX^3} + \beta\frac{d^2\varphi(X)}{dX^2} - \omega^2\varphi(X) = 0 \tag{47}$$

$$\begin{aligned}
&\frac{d^3\varphi(X)}{dX^3} + \frac{1}{2}\alpha\frac{d^2\varphi(X)}{dX^2} = 0 \qquad &&\text{or } \varphi(X) = 0 \\
&\frac{d^2\varphi(X)}{dX^2} = 0 &&\text{or } \frac{d\varphi(X)}{dX} = 0
\end{aligned} \tag{48}$$

A general solution of equation (47) is obtained as follows:

$$\varphi(X) = C_1\exp(\lambda_1 X) + C_2\exp(\lambda_2 X) + C_3\exp(\lambda_3 X) + C_4\exp(\lambda_4 X) \tag{49}$$

where $\lambda_i$ are the roots of the following characteristics equation.

$$\lambda^4 + \alpha\lambda^3 + \beta\lambda^2 - \omega^2 = 0 \tag{50}$$

The constants $C_i$ along with the natural frequencies are determined for different boundary conditions. For **cantilever FG beams**, the following boundary conditions are used:

$$\varphi(0) = 0, \frac{d\varphi(0)}{dX} = 0, \frac{d^2\varphi(1)}{dX^2} = 0, \frac{d^3\varphi(1)}{dX^3} = 0 \tag{51}$$

Thus, the nondimensional natural frequencies can be obtained by solving the following equation:

$$\begin{vmatrix}
1 & 1 & 1 & 1 \\
\lambda_1 & \lambda_2 & \lambda_3 & \lambda_4 \\
\lambda_1^2\exp(\lambda_1) & \lambda_2^2\exp(\lambda_2) & \lambda_3^2\exp(\lambda_3) & \lambda_4^2\exp(\lambda_4) \\
\lambda_1^3\exp(\lambda_1) & \lambda_2^3\exp(\lambda_2) & \lambda_3^3\exp(\lambda_3) & \lambda_4^3\exp(\lambda_4)
\end{vmatrix} = 0 \tag{52}$$

and the constants, $C_i$, are determined as follows:

$$\begin{aligned}
&C_3 = -C_1\frac{\xi_1}{\xi_2} \\
&C_4 = \frac{1}{\lambda_4 - \lambda_2}(-C_1\lambda_1 + \lambda_2(C_1 + C_3) - C_3\lambda_3) \\
&C_2 = -C_1 - C_3 - C_4 \\
&\xi_1 = \lambda_1^2\exp(\lambda_1) - \lambda_2^2\exp(\lambda_2) + \left(\frac{\lambda_1 - \lambda_2}{\lambda_4 - \lambda_2}\right)(\lambda_2^2\exp(\lambda_2) - \lambda_4^2\exp(\lambda_4)) \\
&\xi_2 = \lambda_3^2\exp(\lambda_3) - \lambda_2^2\exp(\lambda_2) + \left(\frac{\lambda_3 - \lambda_2}{\lambda_4 - \lambda_2}\right)(\lambda_2^2\exp(\lambda_2) - \lambda_4^2\exp(\lambda_4))
\end{aligned} \tag{53}$$

As for **simple supported FG beams**, the following boundary conditions are used:

$$\varphi(0) = 0, \frac{d^2\varphi(0)}{dX^2} = 0, \varphi(1) = 0, \frac{d^2\varphi(1)}{dX^2} = 0 \tag{54}$$

The nondimensional natural frequencies are, then, determined by solving the following equation:

$$\begin{vmatrix} 1 & 1 & 1 & 1 \\ \lambda_1^2 & \lambda_2^2 & \lambda_3^2 & \lambda_4^2 \\ \exp(\lambda_1) & \exp(\lambda_2) & \exp(\lambda_3) & \exp(\lambda_4) \\ \lambda_1^2 \exp(\lambda_1) & \lambda_2^2 \exp(\lambda_2) & \lambda_3^2 \exp(\lambda_3) & \lambda_4^2 \exp(\lambda_4) \end{vmatrix} = 0 \qquad (55)$$

and the constants, $C_i$, are obtained as follows:

$$C_3 = -C_1 \frac{\xi_1}{\xi_2}$$

$$C_4 = \frac{1}{\lambda_4^2 - \lambda_2^2}(-C_1\lambda_1^2 + \lambda_2^2(C_1 + C_3) - C_3\lambda_3^2)$$

$$C_2 = -C_1 - C_3 - C_4 \qquad (56)$$

$$\xi_1 = \exp(\lambda_1) - \exp(\lambda_2) + \left(\frac{\lambda_1^2 - \lambda_2^2}{\lambda_4^2 - \lambda_2^2}\right)(\exp(\lambda_2) - \exp(\lambda_4))$$

$$\xi_2 = \exp(\lambda_3) - \exp(\lambda_2) + \left(\frac{\lambda_3^2 - \lambda_2^2}{\lambda_4^2 - \lambda_2^2}\right)(\exp(\lambda_2) - \exp(\lambda_4))$$

For **clamped-clamped FG beams**, the following boundary conditions are employed:

$$\varphi(0) = 0 \; , \; \varphi(1) = 0 \; , \frac{d\varphi(0)}{dX} = 0 \; , \frac{d\varphi(1)}{dX} = 0 \qquad (57)$$

Therefore, the nondimensional natural frequencies can be obtained by solving the following equation:

$$\begin{vmatrix} 1 & 1 & 1 & 1 \\ \lambda_1 & \lambda_2 & \lambda_3 & \lambda_4 \\ \exp(\lambda_1) & \exp(\lambda_2) & \exp(\lambda_3) & \exp(\lambda_4) \\ \lambda_1 \exp(\lambda_1) & \lambda_2 \exp(\lambda_2) & \lambda_3 \exp(\lambda_3) & \lambda_4 \exp(\lambda_4) \end{vmatrix} = 0 \qquad (58)$$

also, the constants, $C_i$, are determined as follows:

$$C_3 = -C_1 \frac{\xi_1}{\xi_2}$$

$$C_4 = \frac{1}{\lambda_4 - \lambda_2}(-C_1\lambda_1 + \lambda_2(C_1 + C_3) - C_3\lambda_3)$$

$$C_2 = -C_1 - C_3 - C_4 \qquad (59)$$

$$\xi_1 = \exp(\lambda_1) - \exp(\lambda_2) + \left(\frac{\lambda_1 - \lambda_2}{\lambda_4 - \lambda_2}\right)(\exp(\lambda_2) - \exp(\lambda_4))$$

$$\xi_2 = \exp(\lambda_3) - \exp(\lambda_2) + \left(\frac{\lambda_3 - \lambda_2}{\lambda_4 - \lambda_2}\right)(\exp(\lambda_2) - \exp(\lambda_4))$$

## 5.   Effects of surface integrity on initial curvatures, frequencies and mode shapes of FG beams

In this section, effects of the surface integrity on the initial curvatures, natural frequencies, and mode shapes of FG nanobeams with different boundary conditions are discussed. To this end, a set of FG nanobeams with the geometrical and material characteristics presented in Table 1 are studied.

### 5.1. Initial curvatures of FG nanobeams

Figures 2-4 show the initial deflections produced in unloaded cantilever, simple supported, and clamped-clamped FG beams due to surface residual stresses. As presented in equations (40) and (41), FG beams are subjected to surface stresses-based forces and/or moments. These residual forces and moments mainly depend on the difference between the surface stresses of the upper and lower surfaces of a FG beam. Moreover, the magnitudes of these residual forces and moments depend on the material grading parameter, $n$. As observed in Figures 2-4, an increase in the material grading parameter is accompanied with a decrease in the nondimensional beam deflection.

Figures 2-4 reflect the significant influence of the surface texture on the initial curvature of FG nanobeams. Thus, a FG beam exhibits an increase in its initial curvature for an increase in the slope of its surface roughness. When compared to Gurtin-Murdoch surface elasticity model ($R_a \rightarrow 0$, $W_a \rightarrow 0$, $RS \rightarrow 0$, $WS \rightarrow 0$), the surface integrity model accounts for the coupled effects of the surface energy and surface texture. It is clear that different deflections are obtained when modeling FG beams using Gurtin-Murdoch model and the surface integrity model. Modeling FG beams using Gurtin-Murdoch model may result in under/over-estimations of the beam initial curvature. Therefore, to accurately detect the initial curvatures of FG nanobeams due to residual stresses of their surfaces, the surface integrity model is preferred over Gurtin-Murdoch surface model.

Table 1: Geometrical and material properties of the studied FG nanobeams.

| Geometrical properties | |
|---|---|
| Beam thickness, $h$ | $1\,nm \rightarrow 15\,nm$ |
| Beam width, $b$ | $2h$ |
| Beam length, $L$ | $50h$ |
| **Surface texture** | |
| Average surface roughness, $R_a$ | $0.1\,nm \rightarrow 10\,nm$ |
| Average surface waviness, $W_a$ | $0.1\,nm$ |
| Average slope of surface roughness, $RS$ | $1\,nm/L \rightarrow 10\,nm/L$ |
| Average slope of surface waviness, $WS$ | $0.02\,RS$ |
| **Material constituents** | |
| ***Upper surface (silicon)*** | |
| Equivalent elastic modulus, $E^+$ | $169\,GPa$ |
| Equivalent surface elastic modulus, $E_S^+$ | $-10.036\,N/m$ |
| Surface residual stress, $\sigma_S^+$ | $0.605\,N/m$ |
| ***Lower surface (aluminum)*** | |
| Equivalent elastic modulus, $E^-$ | $90.24\,GPa$ |
| Equivalent surface elastic modulus, $E_S^-$ | $6.0905\,N/m$ |
| Surface residual stress, $\sigma_S^-$ | $0.91\,N/m$ |
| Material grading parameter, $n$ | $0 \rightarrow \infty$ |

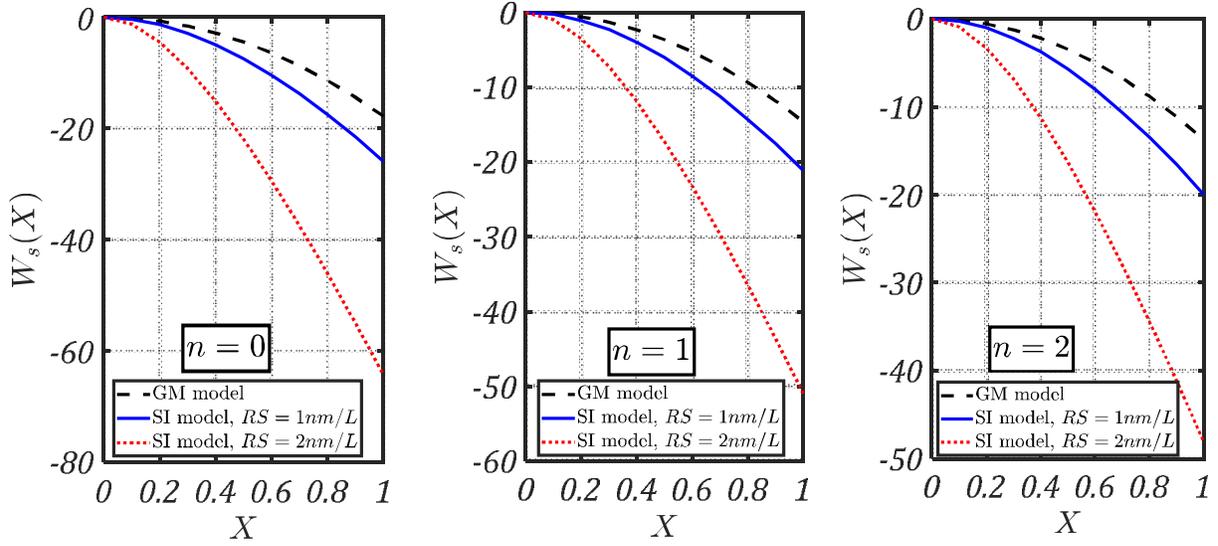

Figure 2: Initial curvatures of **cantilever** FG nanobeams due to surface residual stresses: comparison between surface integrity (SI) model and Gurtin-Murdoch (GM) surface model ($R_a = 0.1nm$, $WS = 0.02RS$, $W_a = 0.1nm$, $h = 5nm$).

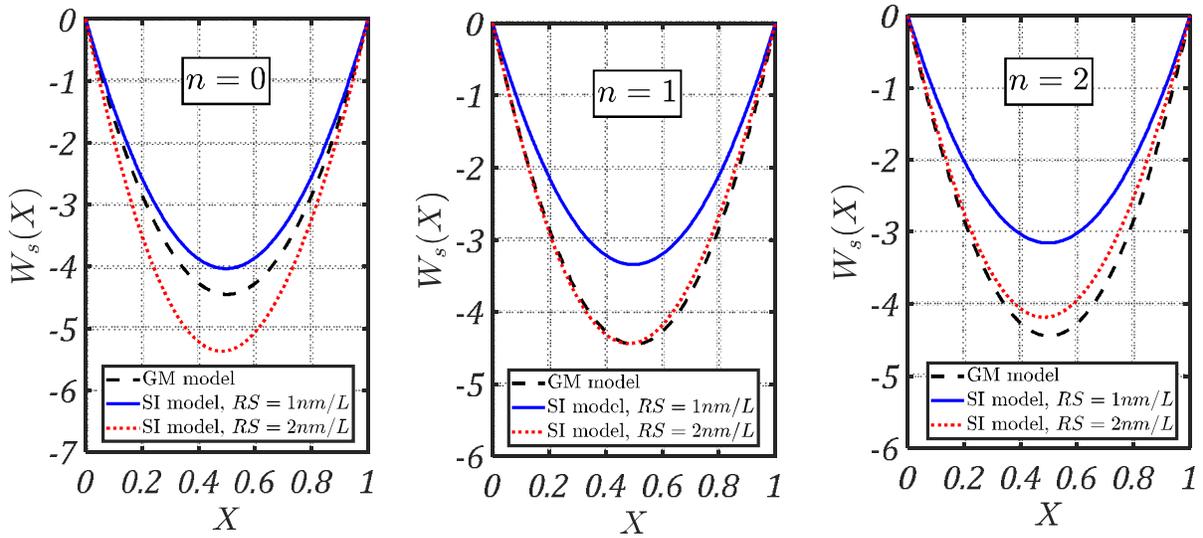

Figure 3: Initial curvatures of **simple supported** FG nanobeams due to surface residual stresses: comparison between surface integrity (SI) model and Gurtin-Murdoch (GM) surface model ($R_a = 0.1nm$, $WS = 0.02RS$, $W_a = 0.1nm$, $h = 5nm$).

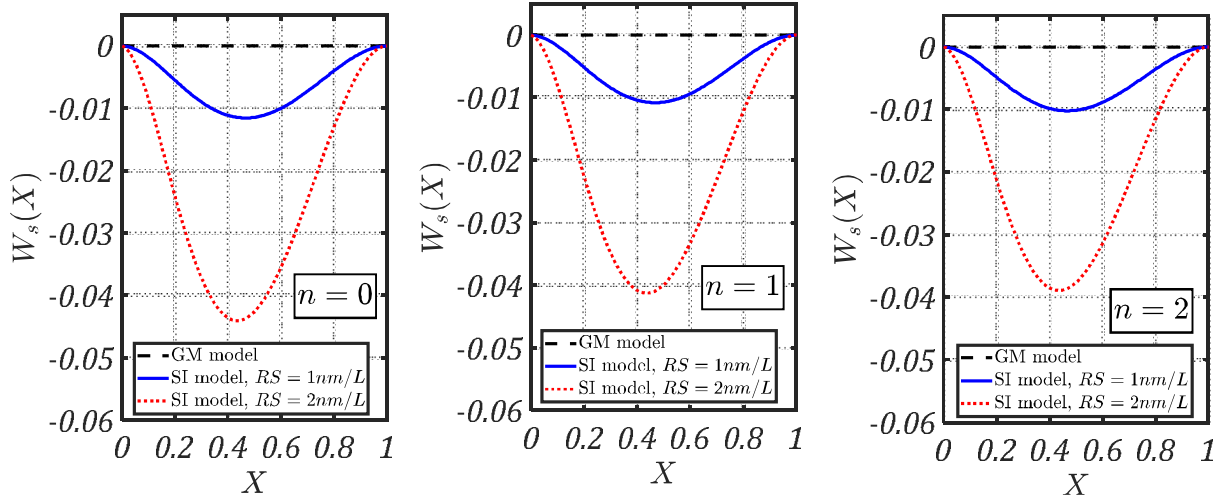

Figure 4: Initial curvatures of **clamped-clamped** FG nanobeams due to surface residual stresses: comparison between surface integrity (SI) model and Gurtin-Murdoch (GM) surface model ($R_a = 0.1nm$, $WS = 0.02RS$, $W_a = 0.1nm$, $h = 5nm$).

### 5.2. Influence of surface integrity on natural frequencies of FG nanobeams

To show the influence of the surface integrity on the natural frequencies of FG nanobeams, the first four nondimensional natural frequencies ($\omega_n = \Omega_n \sqrt{I_0 L^4 / E^+ + \left(\frac{bh^3}{12}\right)}$ where $\Omega_n$ is the dimensional natural frequency) are presented as functions of the average slope of the surface roughness (Figures 5-7) and functions of the average surface roughness (Figures 8-10) for different values of the beam thickness and for the different boundary conditions.

#### 5.2.1 Effects of the average slope of surface roughness

Figure 5 shows effects of the surface roughness on the nondimensional natural frequencies of cantilever FG beams. As shown in the figure, the first two nondimensional natural frequencies decrease with an increase in the average slope of the surface roughness, $RS$. Thus, depending on the beam size, the natural frequency of a cantilever FG beam may decrease to zero resulting in a zero-frequency mode. In contrast, the nondimensional natural frequencies of the third or higher modes increase with an increase in the average slope of the surface roughness.

Figure 6 shows the influence of the surface roughness on the nondimensional natural frequencies of simple supported FG nanobeams. For simple supported FG nanobeams, the nondimensional natural frequencies of the higher-order modes increase with an increase in the average slope of the surface roughness. The fundamental natural frequency, however, increases then decreases with an increase in the average slope of the surface roughness.

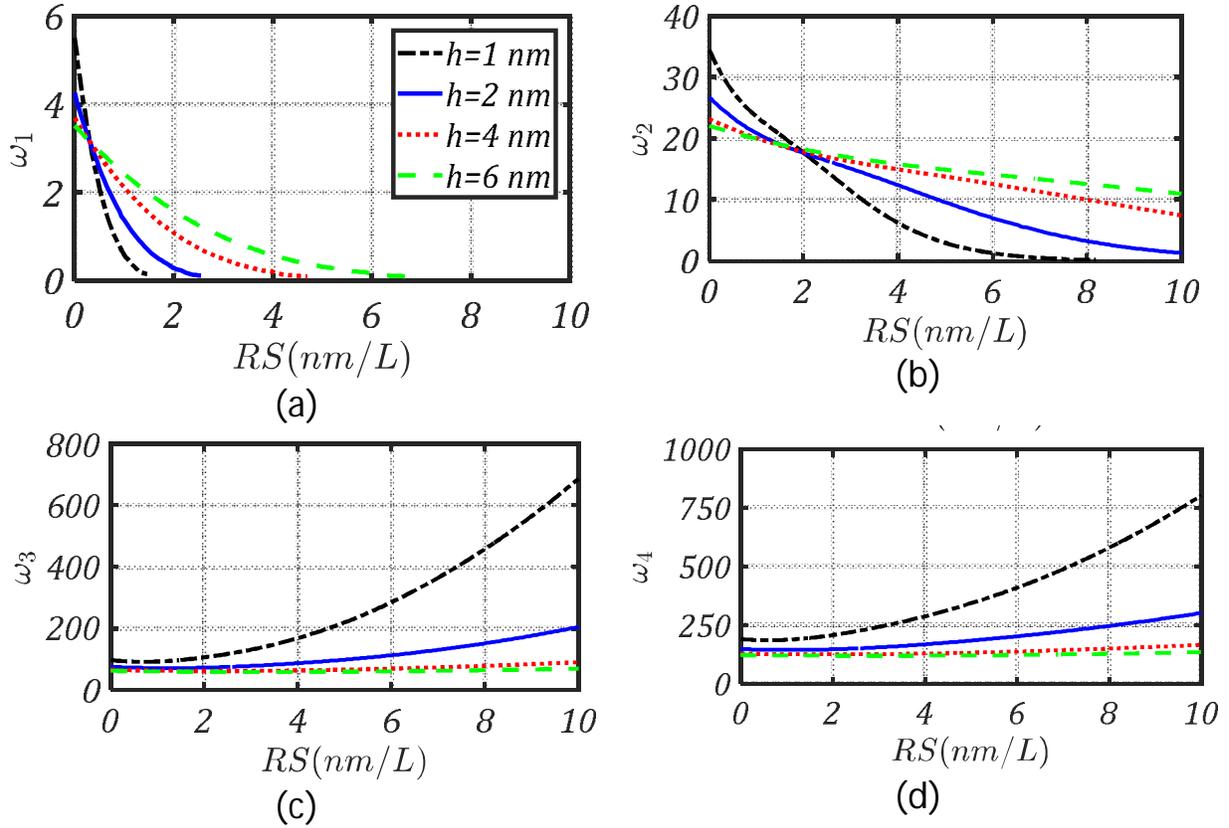

Figure 5: The first four nondimensional natural frequencies ($\omega_n = \Omega_n \sqrt{I_0 L^4 / E^+ \left( \frac{bh^3}{12} \right)}$) as functions of the average slope of the surface roughness, $RS$, of **cantilever** FG nanobeams ($R_a = 0.1nm$, $WS = 0.02RS$, $W_a = 0.1nm$, $n = 2$).

The nondimensional natural frequencies of clamped-clamped FG nanobeams are plotted versus the average slope of the surface roughness in Figure 7. Unlike cantilever and simple supported nanobeams, all the nondimensional natural frequencies of clamped-clamped FG beams increase with an increase in the average slope of the surface roughness.

It follows from the presented results in Figures 5-7 that the changes in the natural frequencies of FG nanobeams due to an increase in the average slope of the surface roughness depend on the boundary conditions. Thus, FG beams with different boundary conditions exhibit different vibrational behaviors. An interpretation of the foregoing observations is given in Section 5.3 by discussing the mode localization phenomenon of FG beams due surface textures. Moreover, it can observed in Figures 5-7 that the rate of change in the nondimensional natural frequencies increases with a decrease in the beam size. This indicates that the contribution of the surface integrity increases with a decrease in the beam size. The results presented in Figures 5-7 indicate that surface integrity strongly affect the vibrational behaviors of FG nanobeams, and special considerations should be given for the slope of the surface roughness when investigating the frequencies of FG nanobeams.

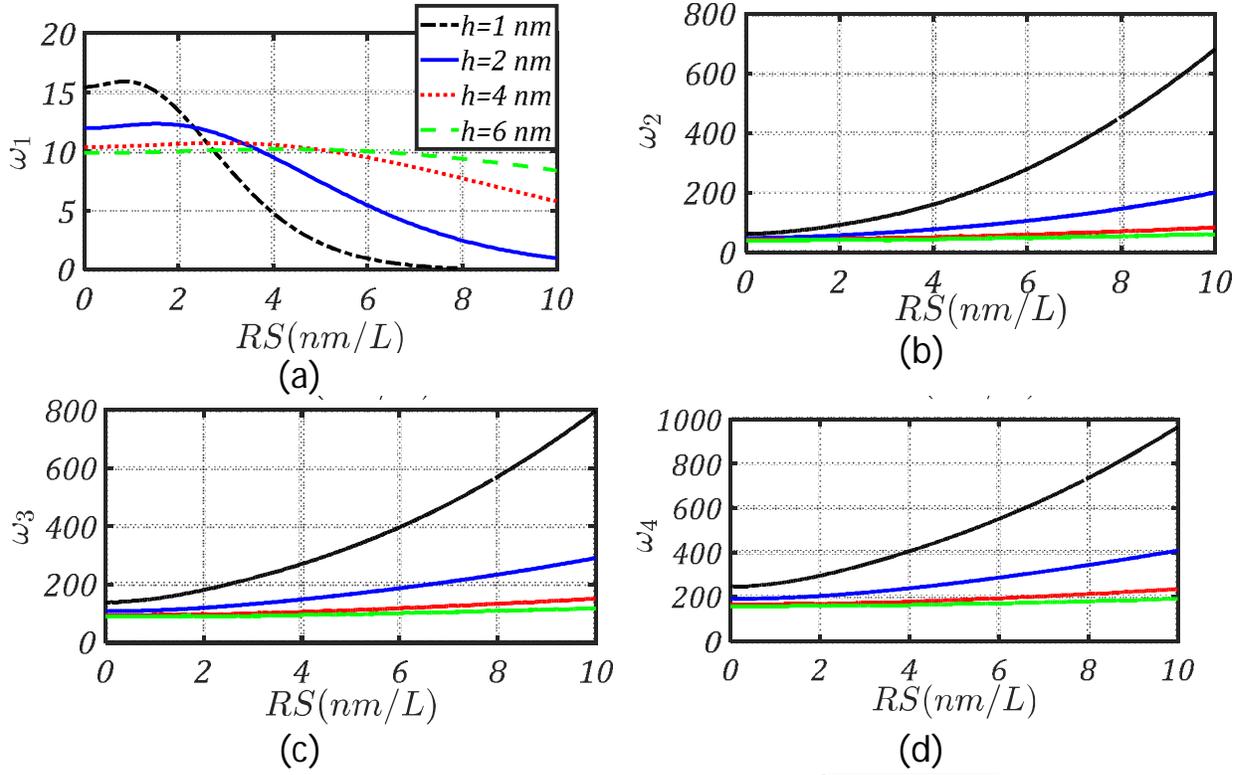

Figure 6: The first four nondimensional natural frequencies ($\omega_n = \Omega_n \sqrt{I_0 L^4 / E^+ \left(\frac{bh^3}{12}\right)}$) as functions of the average slope of the surface roughness, $RS$, of **simple supported** FG nanobeams ($R_a = 0.1nm$, $WS = 0.02RS$, $W_a = 0.1nm$, $n = 2$).

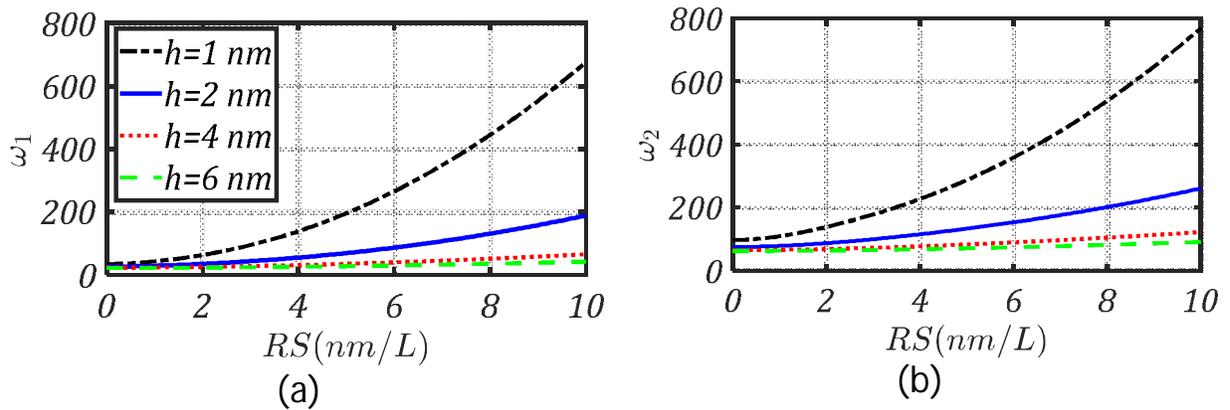

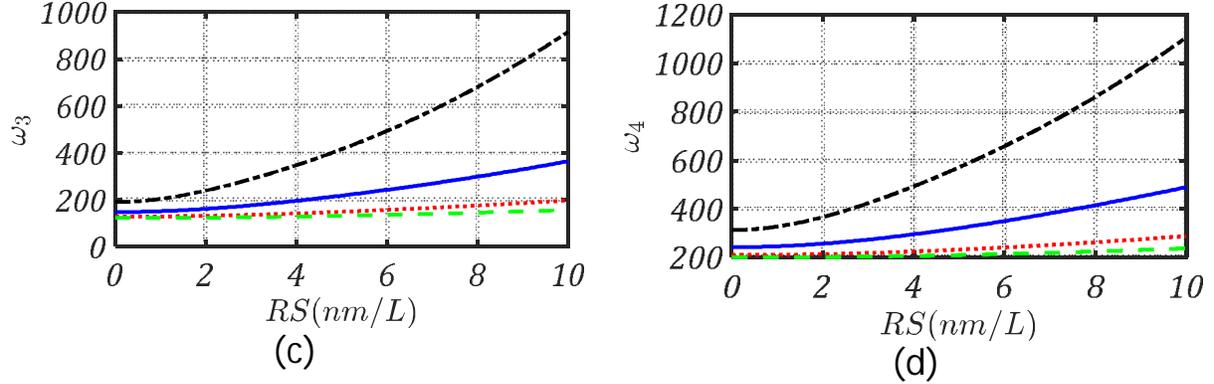

Figure 7: The first four nondimensional natural frequencies ($\omega_n = \Omega_n \sqrt{I_0 L^4 / E^+ \left(\frac{bh^3}{12}\right)}$) as functions of the average slope of the surface roughness, $RS$, of **clamped-clamped** FG nanobeams ($R_a = 0.1nm$, $WS = 0.02RS$, $W_a = 0.1nm$, $n = 2$).

### 5.2.2   Effects of the average surface roughness

Figures 8-10 reveal the influence of the average surface roughness, $R_a$, on the natural frequencies of FG nanobeams. It follows from these figures that an increase in the average surface roughness is accompanied with an increase in the nondimensional natural frequencies of FG beams with the different boundary conditions. Moreover, it is clear that the rate of increase in the natural frequency of FG nanobeams due to the average surface roughness increases with a decrease in the beam size. In can be observed from Figures 8 and 9 that cantilever and simple supported FG beams with small values of the average surface roughness may exhibit zero-frequency modes. Thus, with the decrease in the average surface roughness, the first two natural frequencies of a cantilever FG beam are observed decreasing to zeros, as shown in Figure 8. On the other hand, only the fundamental natural frequency of a simple supported FG beam deceases to zero with a decrease in the average surface roughness. In contrast to cantilever and simple supported FG beams, clamped-clamped FG beams does not reflect any zero-frequency modes for the change in the average surface roughness.

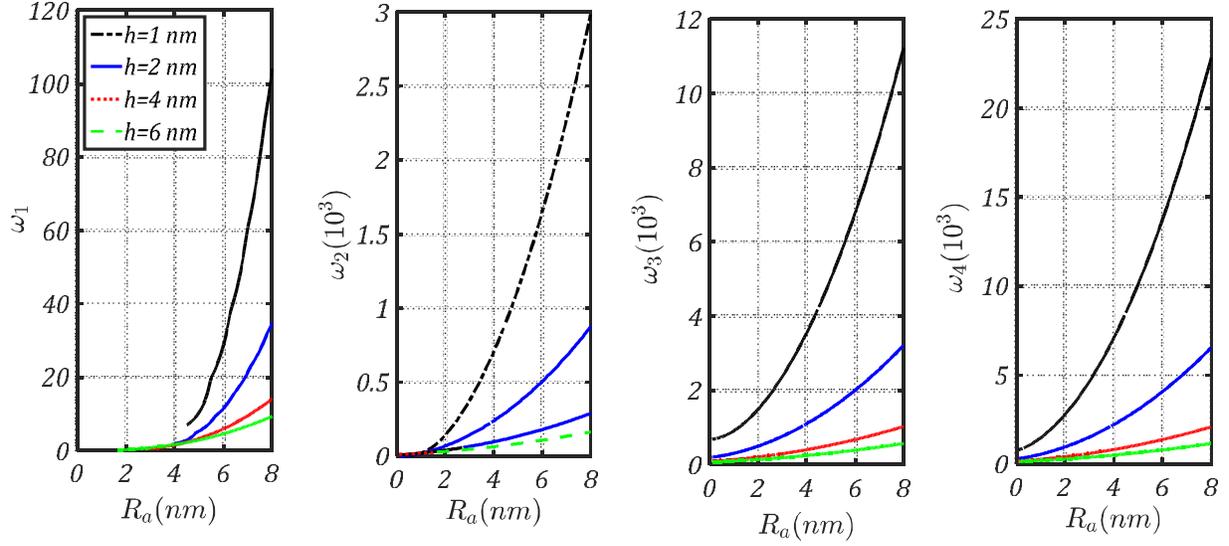

Figure 8: The first four nondimensional natural frequencies ($\omega_n = \Omega_n\sqrt{I_0 L^4/E^+\left(\frac{bh^3}{12}\right)}$) as functions of the average surface roughness, $R_a$, of **cantilever** FG nanobeams ($RS = 10nm/L$, $WS = 0.02RS$, $W_a = 0.1nm$, $n = 2$).

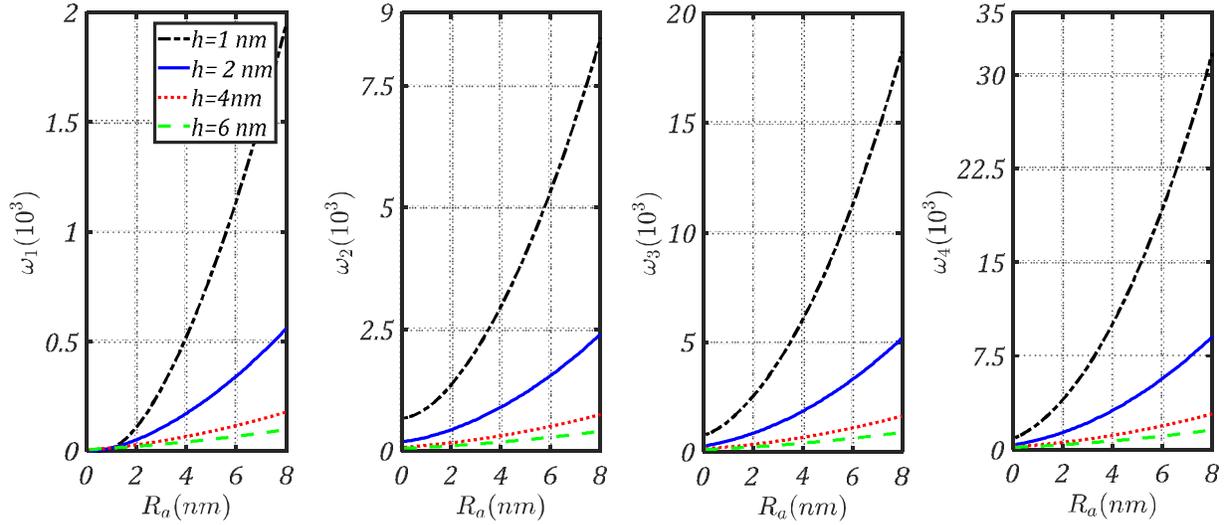

Figure 9: The first four nondimensional natural frequencies ($\omega_n = \Omega_n\sqrt{I_0 L^4/E^+\left(\frac{bh^3}{12}\right)}$) as functions of the average surface roughness, $R_a$, of **simple supported** FG nanobeams ($RS = 10nm/L$, $WS = 0.02RS$, $W_a = 0.1nm$, $n = 2$).

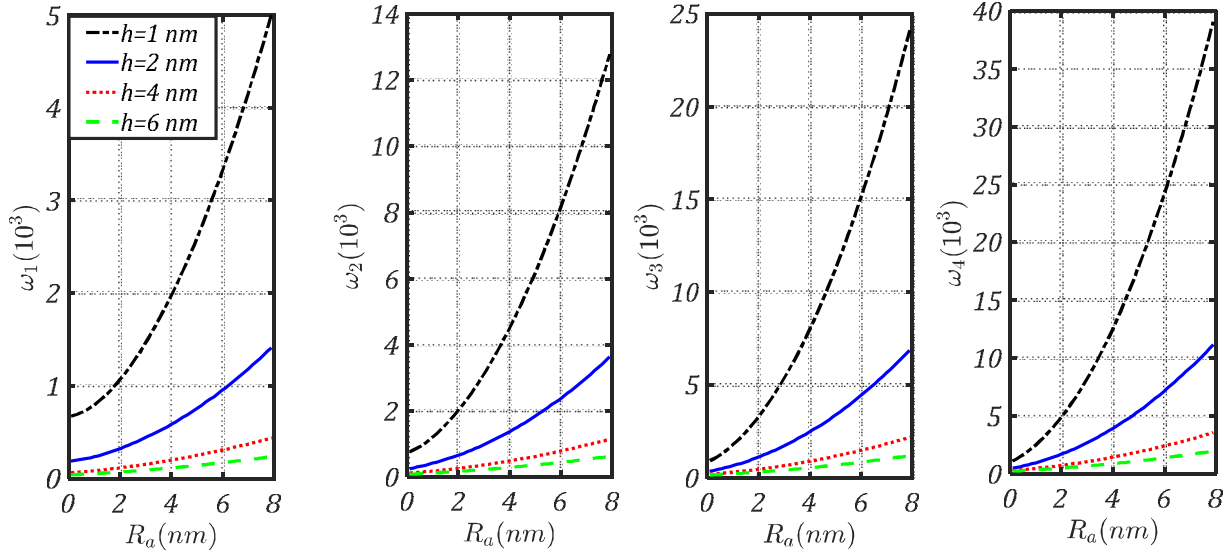

Figure 10: The first four nondimensional natural frequencies ($\omega_n = \Omega_n \sqrt{I_0 L^4 / E^+ \left(\frac{bh^3}{12}\right)}$) as functions of the average surface roughness, $R_a$, of **clamped-clamped** FG nanobeams ($RS = 10nm/L$, $WS = 0.02RS$, $W_a = 0.1nm$, $n = 2$).

### 5.3. Mode localization of FG nanobeams due to surface integrity

Figures 11-13 and Figures 14-16 show the influence of the average slope of the surface roughness, $RS$, and the average surface roughness, $R_a$, on the first four mode shapes of FG nanobeams with different boundary conditions. These figures are used to portray the dependency of the mode shapes of FG nanobeams on the surface roughness. The mode shapes are depicted for different values of the average slope of the surface roughness, $RS$, and the average surface roughness, $R_a$, portraying the evolution of the mode localization of FG beams.

Figure 11 shows the mode shapes of a cantilever FG nanobeam. Different characteristics of the mode shape can be observed. As shown in Figures 11(a) and 11(b), the amplitudes of vibration of the first two modes increase with an increase in the average slope of the surface roughness up to $RS = 1.45\ nm/L$ and $RS = 8.1\ nm/L$, respectively. Beyond the aforementioned values of the average slope of roughness, the first two modes are zero-frequency modes. The first two modes are nonlocalized modes where the vibration energy is distributed over the whole beam length, as shown in Figures 11(a) and 11(b). Moreover, the increase in the amplitude of the first two modes indicates that the surface roughness allows vibration energy to propagate and cover longer spans of the beam and, hence, a decrease in the natural frequencies (the result which was previously demonstrated in Figures 5 (a) and 5(b)).

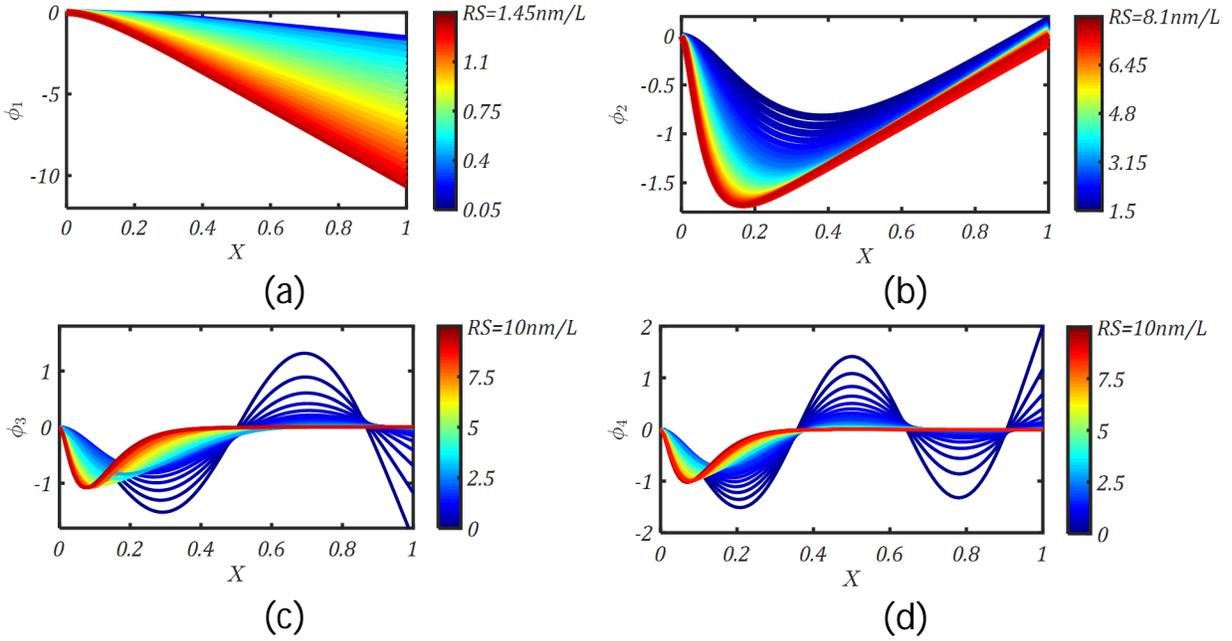

Figure 11: Mode localization phenomenon of **cantilever** FG nanobeams due to an increase in the average slope of the surface roughness, $RS$ ($R_a = 0.1nm/L$, $WS = 0.02RS$, $W_a = 0.1nm$, $h = 1nm$, $n = 2$).

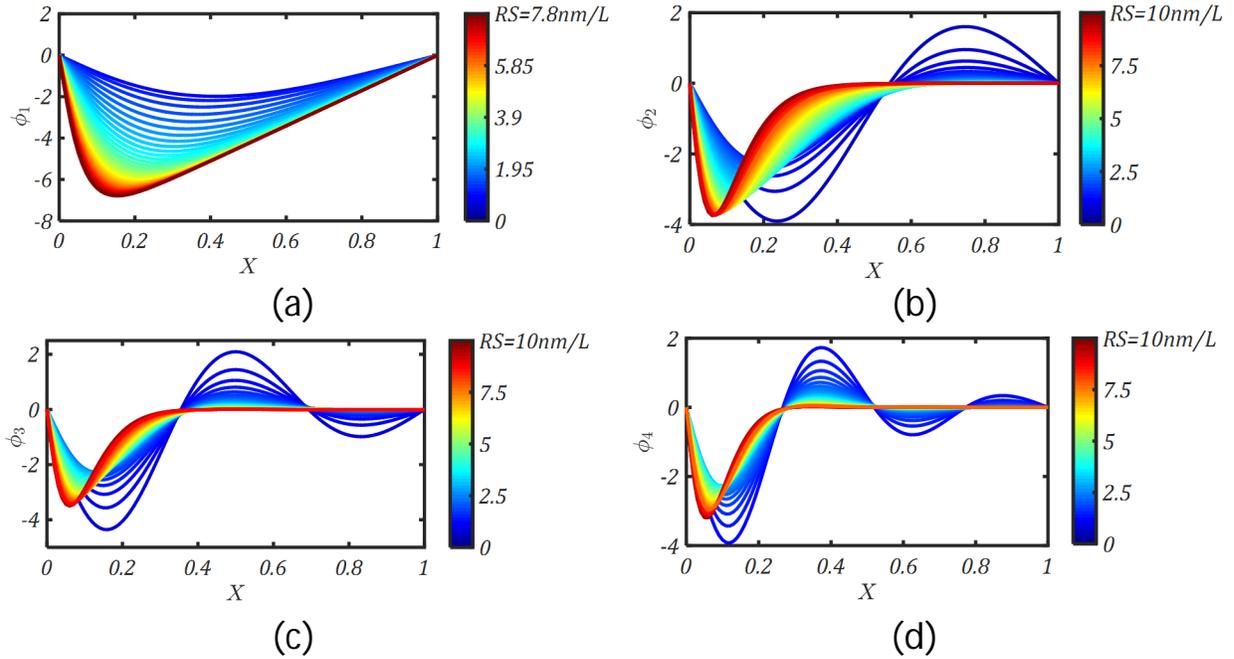

Figure 12: Mode localization phenomenon of **simple supported** FG nanobeams due to an increase in the average slope of the surface roughness, $RS$ ($R_a = 0.1nm/L$, $WS = 0.02RS$, $W_a = 0.1nm$, $h = 1nm$, $n = 2$).

Figures 11(c) and 11(d) portray the localization of the 3rd and 4th modes of vibration of cantilever FG beams due to an increase in the average slope of the surface roughness. It is clear that the surface roughness inhibits the propagation of the vibration energy of these higher-order modes throughout the beam length. The inhibition of the vibration energy increases with an increase in the average slope of the surface roughness. As a result, the magnitude of vibration decreases, and the vibration energy is confined within a small portion of the beam causing a mode localization. These observations explain the increase in the natural frequency of the 3rd and 4th modes previously seen in Figures 5(c) and 5(d).

Figure 12 shows the first four mode shapes of simple supported FG beams with different values of the average slope of the surface roughness. As shown in Figure 12(a), no mode localization is observed for the fundamental mode shape where the vibration energy is distributed over the whole beam length. Inspecting Figure 12(a), it follows that the magnitude of the fundamental mode decreases with an increase in the average slope of the surface roughness up to $RS \cong 1.5 \; nm/L$. Then, the magnitude increases with an increase in the average slope of the roughness up to $R_S = 7.8 \; nm/L$. This explains the observed trends of the nondimensional fundamental natural frequency observed in Figure 6(a).

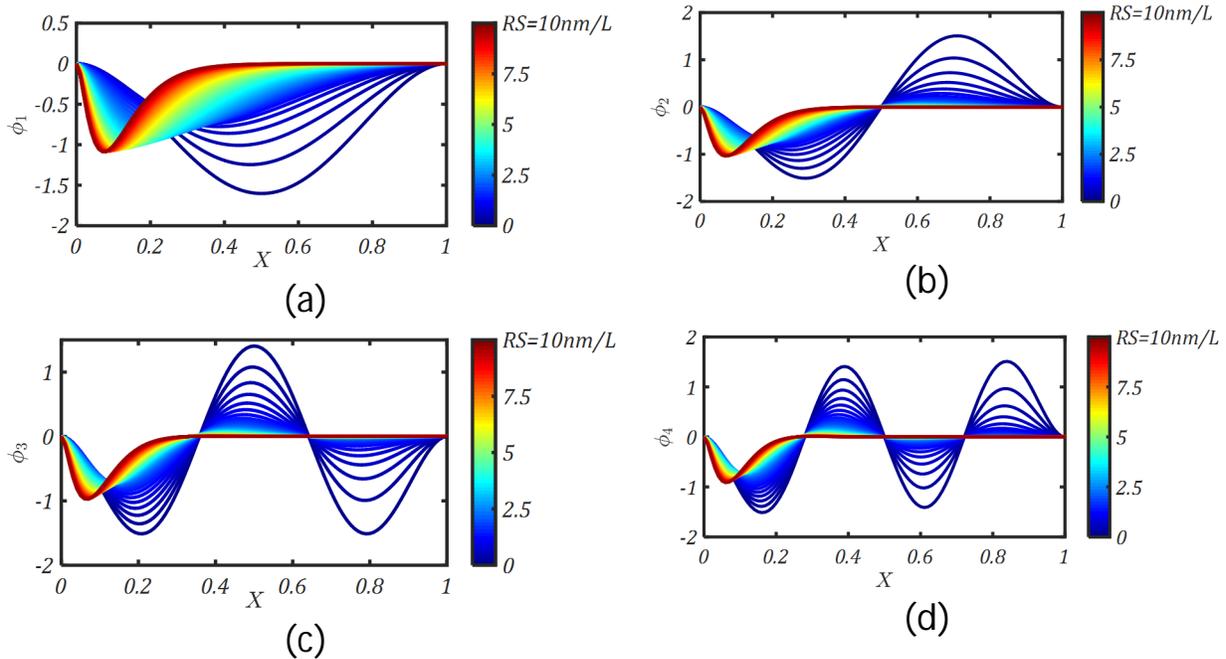

Figure 13: Mode localization phenomenon of **clamped-clamped** FG nanobeams due to an increase in the average slope of the surface roughness, $RS$ ($R_a = 0.1 nm/L$, $WS = 0.02RS$, $W_a = 0.1nm$, $h = 1nm$, $n = 2$).

Figures 12(b)-12(d) show the evolution of the mode localization of simple supported beams with an increase in the average slope of the surface roughness. An increase in the average slope of the surface roughness is accompanied with an increase in the inhibition of the higher-order mode vibrations. As a consequence, higher-order modes strongly localized at high values of the average slope of the surface roughness. The confinement of the vibration energy in a small portion of the beam increases the natural frequency of the beam (as it was previously demonstrated in Figures 6(b)-6(d))

As for clamped-clamped FG beams, all modes are localized due to an increase in the average slope of the surface roughness. Thus, the vibration energy is confined over a small portion of the beam length, as shown in Figure 13. Because of the confinement of the vibration energy over a small portion of the beam, the beam natural frequencies increase. This explains the increase in the natural frequencies of clamped-clamped FG beams observed in Figure 7.

Figures 14-16 show the influence of the average surface roughness, $R_a$, on the first four modes of cantilever, simple supported, and clamped-clamped FG beams, respectively. The presented results in these figures demonstrate that the mode localization of FG beams is mainly due to the slope of the surface roughness, $RS$. Thus, the role of the slope of the surface roughness increases with a decrease in the average roughness, $R_a$. The 3rd and 4th-order modes of cantilever, the higher-order modes of simple supported, and all modes of clamped-clamped FG beams are localized at small value of the average surface roughness. This indicates that the increase in the beam bending stiffness, $D$, due to an increase in the average roughness is higher than the increase in the other stiffnesses, $K$ and $B$. As a consequence, for high values of the average roughness, the contribution of the stiffnesses, $K$ and $B$, to the beam vibration demolishes and hence the vibration energy is distributed over the whole beam length.

It follows from the presented results in Figures 11-16 that a mode localization is conjugate to a natural frequency increase. However, when a frequency decreases, no mode localization takes place due to surface roughness.

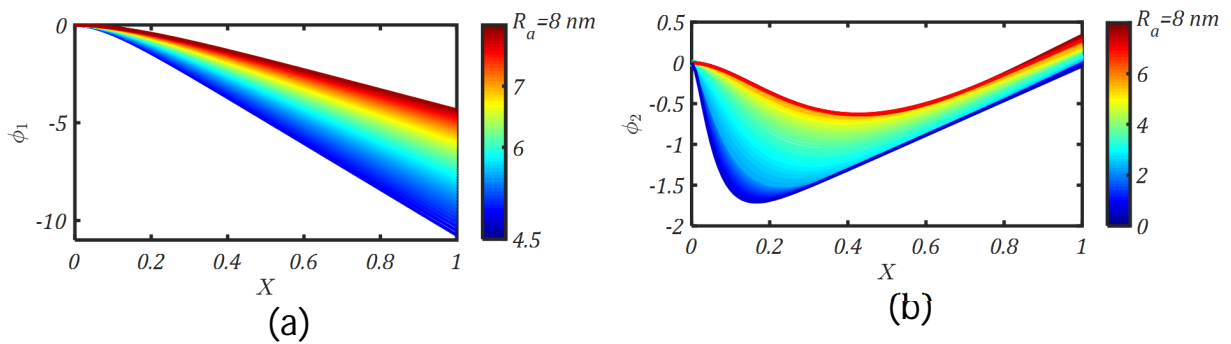

(a) (b)

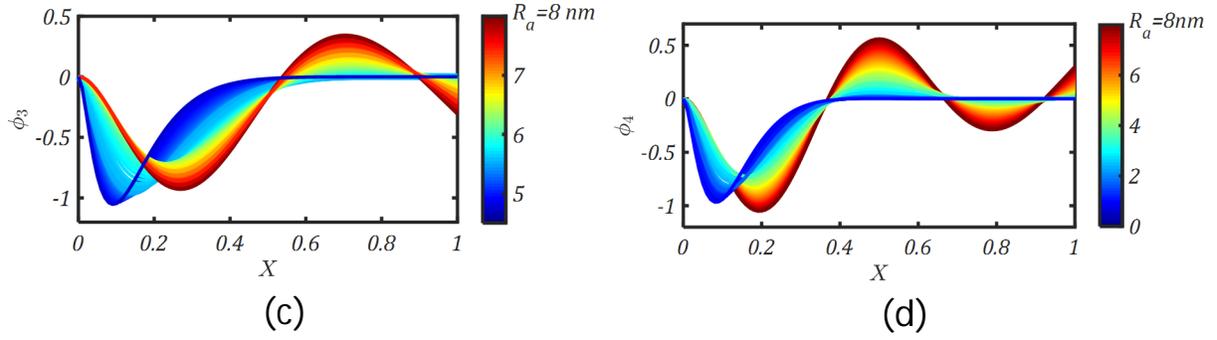

Figure 14: The influence of the average surface roughness, $R_a$, on the first four modes of **cantilever** FG nanobeams ($RS = 10nm/L$, $WS = 0.02RS$, $W_a = 0.1nm$, $h = 1nm$, $n = 2$).

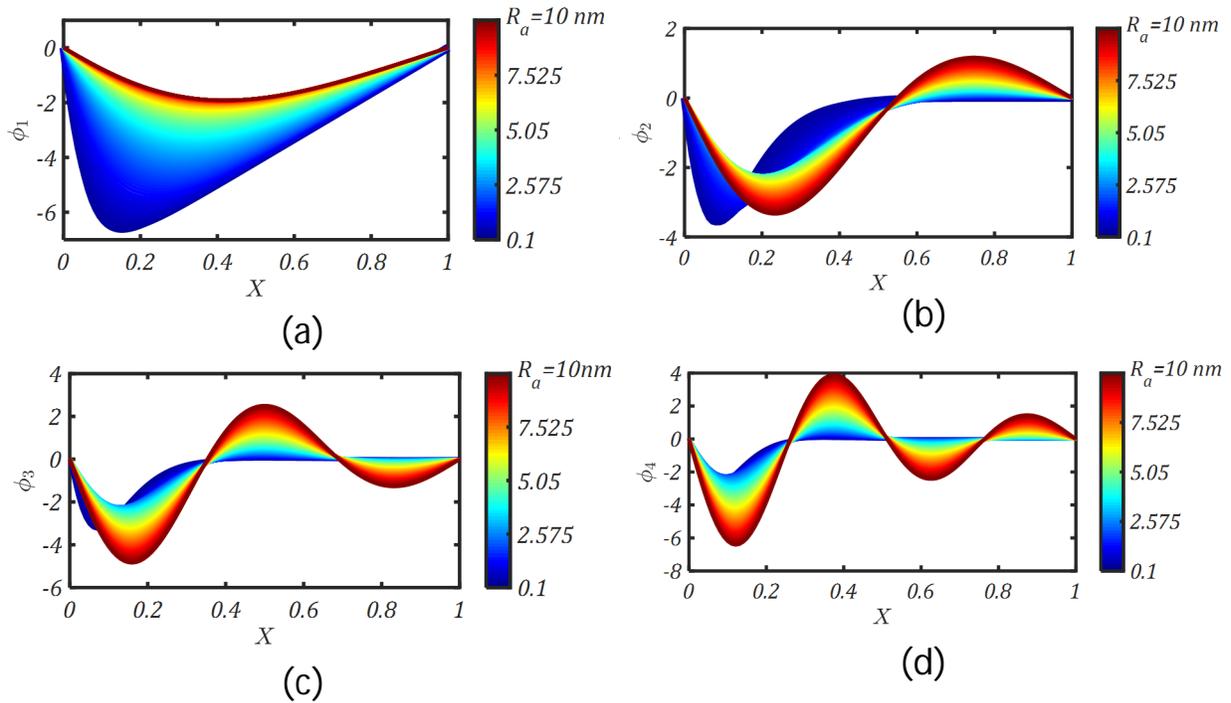

Figure 15: The influence of the average surface roughness, $R_a$, on the first four modes of **simple-supported** FG nanobeams ($RS = 10nm/L$, $WS = 0.02RS$, $W_a = 0.1nm$, $h = 1nm$, $n = 2$).

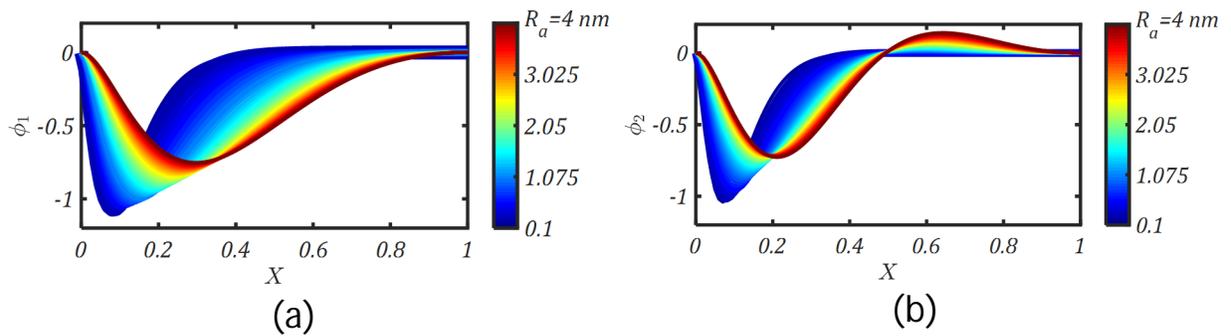

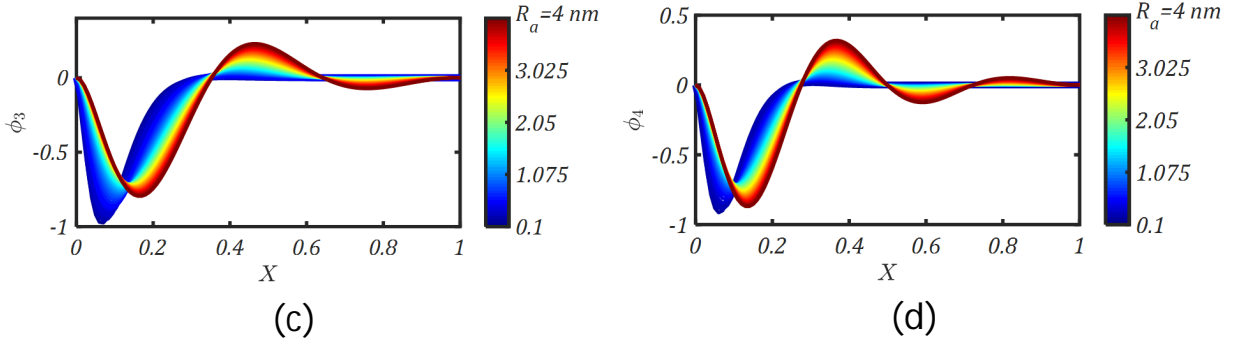

Figure 16: The influence of the average surface roughness, $R_a$, on the first four modes of **clamped-clamped** FG nanobeams ($RS = 10nm/L$, $WS = 0.02RS$, $W_a = 0.1nm$, $h = 1nm$, $n = 2$).

## 6. Surface integrity model versus Gurtin-Murdoch surface model

In this section, a comparison between the results obtained by the surface integrity model and the results obtained using Gurtin-Murdoch model is provided. As previously demonstrated, the surface integrity model outweighs Gurtin-Murdoch surface elasticity model by modeling the coupled effect of the surface texture and the surface excess energy. Thus, the surface integrity model incorporates measures that show the surface energy depends on the surface texture. Gurtin-Murdoch model assumes a nominal surface (*i.e.* smooth surfaces) of the material. However, a real-engineering surface has a texture that is described by the surface waviness and roughness. The surface integrity model incorporates two coupling stiffnesses, $K$ and $B$, which depend on the surface texture and the surface mechanical properties. Moreover, the surface integrity model accounts for the effects of the surface residual stresses where residual forces and moments are generated depending on not only the surface stress but also the surface texture. It should be mentioned that, both surface integrity model and Gurtin-Murdoch model outperform the classical model where they incorporate measures for the surface effects.

### 6.1 Comparison with the classical model

Figures 17-19 show a comparison between the surface integrity model and the classical model. Thus, the nondimensional differences between the natural frequencies predicted by the surface integrity model, $\omega_n$, and the ones predicted by the classical model, $\omega_n^c$, are plotted as functions of the FG beam thickness for different material grading parameters, $n$. The nondimensional natural frequencies of the classical model considered in these analyses are presented in Table 2.

Table 2: The nondimensional natural frequencies of the classical beam model.

|  | Cantilever beam | Simple supported beam | Clamped-clamped beam |
|---|---|---|---|
| $\omega_1^c$ | 3.51602 | 9.86960 | 22.3729 |
| $\omega_2^c$ | 22.03364 | 39.47842 | 61.6728 |
| $\omega_3^c$ | 61.69631 | 88.82644 | 120.9032 |
| $\omega_4^c$ | 120.90102 | 157.91367 | 199.8604 |

For a cantilever FG beam with an average slope of surface roughness $RS = 1nm/L$, the fundamental natural frequency is obtained decreasing with a decrease in the beam thickness (size), as shown in Figure 17. The higher-order natural frequencies, on the other hand, increase with a decrease in the beam size. Figure 18 shows the nondimensional differences between natural frequencies of a simple supported FG beam with an average slope of surface roughness $RS = 6nm/L$ predicted by the surface integrity model and the classical model. For the considered FG beam, the fundamental natural frequency is obtained increasing with a decrease in the beam thickness up to $6\ nm$ due to surface integrity. Then, the trend is switched so that the fundamental natural frequency is observed decreasing with a decrease in the beam thickness lower than $6\ nm$. This behavior matches the previous observations predicted from Figure 6 (a). As for the higher-order modes of the simple supported beam, their natural frequencies increase with a decrease in the beam size. All the natural frequencies of a clamped-clamped FG beam with an average slope of surface roughness $RS = 6nm/L$ are obtained increasing with a decrease in the beam thickness, as shown in Figure 19.

As can be observed in Figures 17-19 the difference between the frequencies obtained by the surface integrity model and the ones of the classical model increases as the beam thickness decreases. This can be attributed to the fact that the contribution of the surface integrity to the mechanics of the FG beam increases as the beam size decreases. Moreover, the figures show the significant effect of the material grading parameter, $n$, where different results are obtained for the different material grading parameters considered in these analyses.

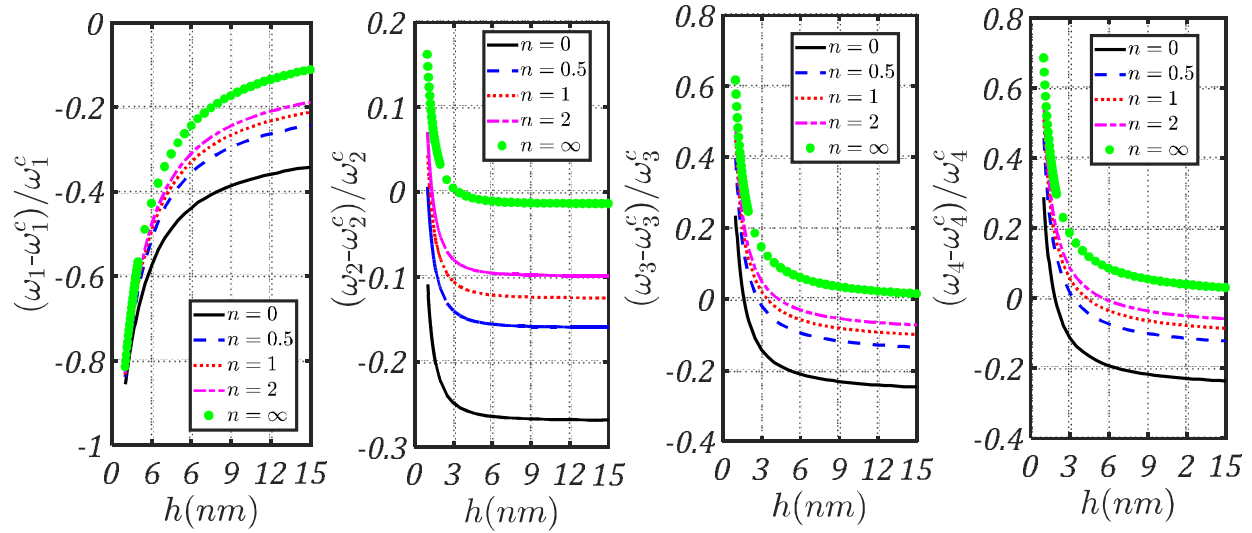

Figure 17: Nondimensional differences between natural frequencies of **cantilever** FG beams predicted by the surface integrity model and the classical model ($RS = 1nm/L$, $R_a = 0.1nm$, $WS = 0.02RS$, $W_a = 0.1nm$).

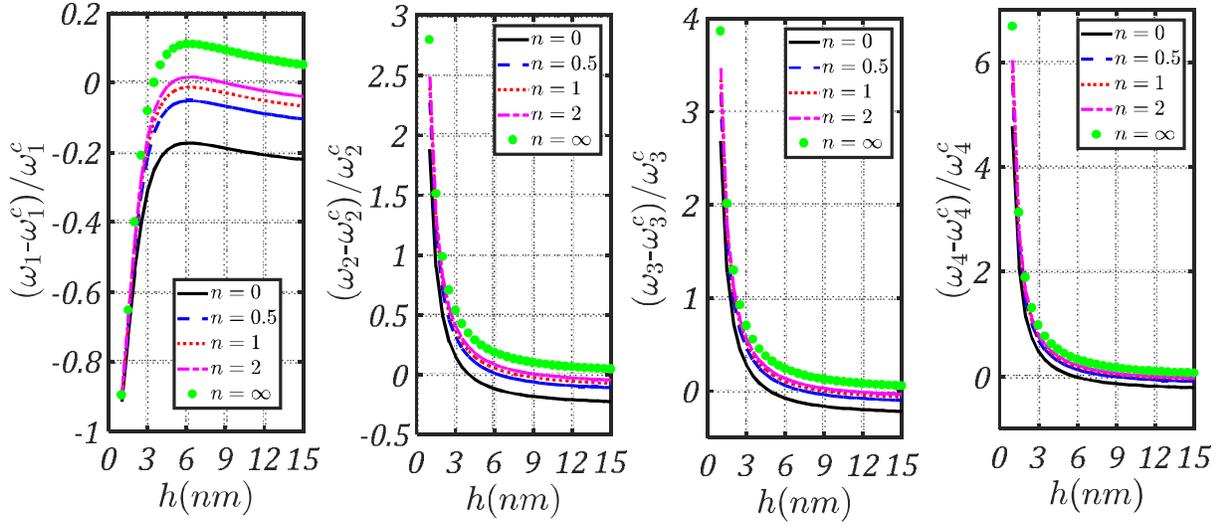

Figure 18: Nondimensional differences between natural frequencies of **simple supported** FG beams predicted by the surface integrity model and the classical model ($RS = 6nm/L$, $R_a = 0.1nm$, $WS = 0.02RS$, $W_a = 0.1nm$).

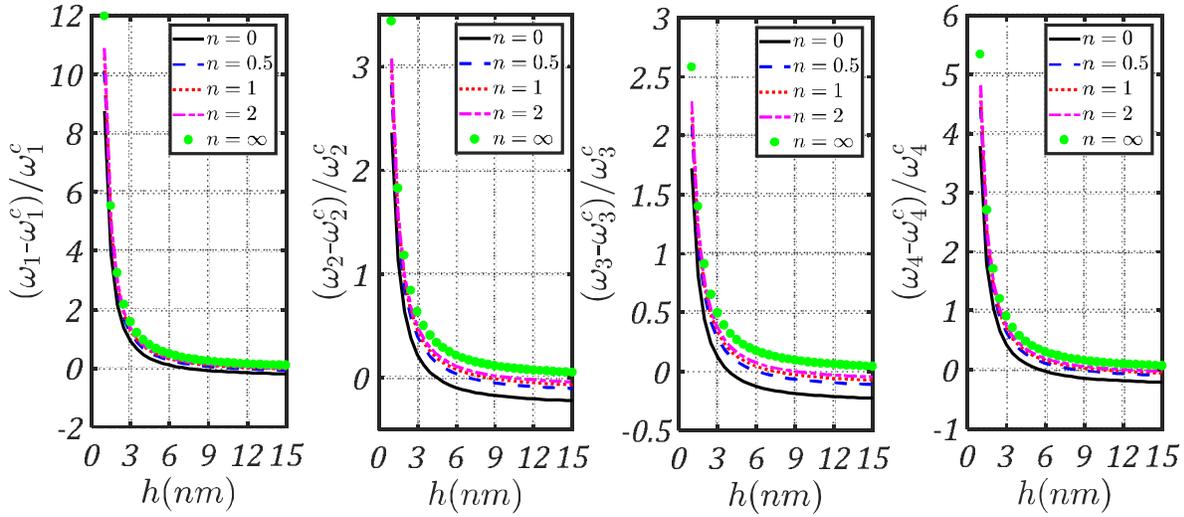

Figure 19: Nondimensional differences between natural frequencies of **Clamped-clamped** FG beams predicted by the surface integrity model and the classical model ($RS = 6nm/L$, $R_a = 0.1nm$, $WS = 0.02RS$, $W_a = 0.1nm$).

## 6.2 Comparison with Gurtin-Murdoch model

Figure 20 presents a comparison between the surface integrity model and Gurtin-Murdoch surface model. The differences between the natural frequencies of the surface integrity model, $\omega_n$, and the ones of Gurtin-Murdoch model, $\omega_n^G$, are presented as functions of the beam thickness for the different boundary conditions. It is clear that a difference between the results of the two models increases as the beam size decreases. This indicates that the incorporation of the surface texture along with the surface excess energy effects is essential and the surface integrity model should be used as the beam size decreases. Gurtin-Murdoch surface model, however, may under/overestimate the natural frequencies of nanobeams.

Generally speaking, the classical model provides wrong estimations of the natural frequencies of nanobeams because it neglects the surface integrity effects. Moreover, because Gurtin-Murdoch model only accounts for the influence of the surface mechanical properties and disregards the surface texture effects, it may give wrong estimations for the surface effects on the mechanics of nanobeams. Thus, the surface integrity model is preferred over the classical model and Gurtin-Murdoch model when investigating the mechanics of nanobeams.

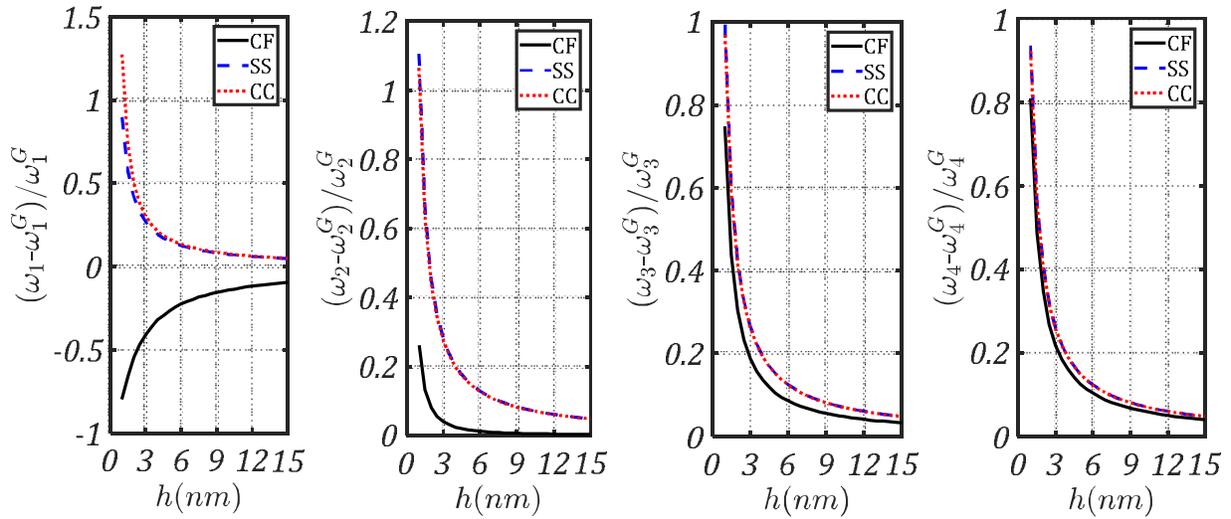

Figure 20: Nondimensional differences between natural frequencies of FG beams predicted by the surface integrity model and Gurtin-Murdoch surface model ($RS = 1nm/L$, $R_a = 0.1nm$, $WS = 0.02RS$, $W_a = 0.1nm$).

## 7. Conclusions

This paper presented the first study on the influence of the surface integrity on the mechanics of FG nanobeams. A detailed formulation for the equation of the motion and the boundary conditions of FG nanobeams with engineering surfaces were derived incorporating measures for the surface texture and surface mechanical properties. Two versions of the proposed surface integrity model were developed. First the model was derived depending on a surface profile function $P(x)$; then, it was reformulated incorporating the average parameters of the surface texture. Analytical solutions for the initial curvatures of cantilever, simple-supported, and clamped-clamped FG beams due surface residual stresses are proposed.

Moreover, the natural frequencies and mode shapes of FG beams with the different boundary conditions were analytically derived.

A parametric study was presented on the influence of the surface integrity on the initial curvatures, frequencies, and mode shapes of FG nanobeams. The initial curvatures of FG beams were observed increasing with an increase in the slope of the surface texture and/or a decrease in the heights of the asperities of the surface roughness. Moreover, it was demonstrated that the natural frequencies of FG beams may decrease or increase due surface integrity depending on the boundary conditions. Thus, the presented results came demonstrating two prospects. It was revealed that surface roughness may lead to a zero-frequency mode or a mode localization. As for the first prospect, surface roughness permits the vibration energy to propagate through the beam length, and hence the mode shapes are nonlocalized. As for this prospect, the surface roughness reduces the beam natural frequency. It was demonstrated that the first two natural frequencies of cantilever FG nanobeams and the fundamental frequency of simple supported FG nanobeams decrease due to surface roughness. Thus, at high values of the average slope of the surface roughness, these frequencies attain zeros. As for the second prospect, surface roughness inhibits the propagation of the vibration energy through the beam length causing a mode localization. Thus, the vibration energy is confined within a small portion of the beam, and natural frequency increases. The higher-order modes of vibration of cantilever and simple supported FG beams as well as all the modes of vibration of clamped-clamped FG beams are localized at a high value of the average slope of the surface roughness.

The proposed surface integrity model for FG nanobeams was compared with Gurtin-Murdoch surface elasticity model and the classical model. The results demonstrated that the surface integrity model outweighs both models where it accounts for, both, surface texture and surface mechanical properties effects. The classical model provides wrong estimations of mechanics of nanobeams because it neglects the influence of the surface integrity. Moreover, because Gurtin-Murdoch model only accounts for the influence of the surface mechanical properties and disregards the surface texture effects, it may give a wrong representation of surface effects on mechanics of nanobeams.